\pdfoutput=1
\documentclass[10pt]{article}
\usepackage{amsmath, bm}
\usepackage{algorithm}
\usepackage{color}
\usepackage{bbm}
\usepackage{epsfig}
\usepackage{amssymb,amsmath,amscd, mdwmath}
\usepackage{graphicx,cite,url}
\usepackage{algorithmic}
\usepackage{float}
\usepackage{amsfonts}
\usepackage{syntonly}
\usepackage{multirow}
\usepackage{graphicx}
\usepackage{subfig}
\usepackage{tabularx}
\usepackage{setspace}
\usepackage{stfloats}
\usepackage{amsthm}
\usepackage[labelfont=bf,labelsep=period,justification=raggedright]{caption}

\makeatletter
\renewcommand{\@biblabel}[1]{\quad#1.}
\makeatother

\date{}

\begin{document}

\begin{flushleft}

{\Large
\textbf{Predicting publication productivity for authors: Shallow or deep architecture?
}

}

Wumei Du, Zheng Xie$^{* }$, Yiqin Lv
\\
    College of Liberal Arts and Sciences, National University of Defense Technology, Changsha, China.
\\  $^*$ xiezheng81@nudt.edu.cn
 \end{flushleft}

\section*{Abstract}

Academic administrators and funding agencies must predict the publication productivity of research groups and individuals to assess authors' abilities. However, such prediction remains an elusive task due to the randomness of individual research and the diversity of authors' productivity patterns. We applied two kinds of approaches to this prediction task: deep neural network learning and model-based approaches. We found that a neural network cannot give a good long-term prediction for groups, while the model-based approaches cannot provide short-term predictions for individuals. We proposed a model that integrates the advantages of both data-driven and model-based approaches, and the effectiveness of this method was validated by applying it to a high-quality dblp dataset, demonstrating that the proposed model outperforms the tested data-driven and model-based approaches.

\noindent {\bf Keywords:}  Scientific publications,  Productivity prediction, Data model.

\section{Introduction}

For the purpose of this study, the publication productivity of an author reflects the number of his or her published scientific papers. This index is associated with promotion, tenure, and individual scientific impact\cite{Garca-Otero2019,Clauset-Larremore2017}. Accordingly, academic administrators and funding agencies must predict this index for research groups and individuals to conduct ability assessments. Consequently, as this index is also related to impact, scientific fields such as scientometrics and informetrics try to establish quantitative analysis methods to predict the scientific impact of authors\cite{Sinatra2016}.

Due to the role of this index in academia, possible factors related to publication productivity, such as academic environment, gender, advisor matching, and mobility across universities, have been discussed\cite{Lindahl-Colliander2019,Ejermo-Fassio2019}. Statistical factor analysis seems to be useful for predicting this index, as it helps address datasets with a large number of observed variables that are thought to reflect a smaller number of latent variables. However, publication productivity results from the behaviors of authors within an open system who are affected by numerous factors; thus, we cannot exhaust all these factors. Moreover, individual research is random, and authors' productivity patterns are diverse\cite{XieLL2018}. Therefore, it is an elusive task to predict publication productivity.

Previous models for predicting scientific performance or impact can be classified into two categories, namely, shallow architectures and deep architectures. Shallow architectures often have fewer parameters than deep architectures; regression models belong to this category. In social systems, numerous factors can affect individual behaviors. However, some of these factors cannot be quantified or measured, and some correlate with each other. It is usually difficult to determine the dominant influencing factors of individual behaviors; thus, shallow architectures often cannot be used to predict the indexes of individual behaviors.

Shallow architectures specifically belong to a certain vertical domain under strict conditions, as the corresponding exogenous variables are typically specific;
on the other hand, deep architectures can be applied under general conditions.
Deep architectures include deep neural networks~(DNNs), recurrent neural networks~(RNNs), and convolutional neural networks~(CNNs)\cite{Schmidhuber2015}.
The adjective ``deep'' comes from the use of multiple layers in those networks. Deep architectures can permit practical applications while retaining their universality under mild conditions. Notably, under mild conditions, the requirement on the information of exogenous variables is diminished; thus, deep architectures can be used to predict individual behaviors. However, the presence of multiple layers means numerous parameters, thereby requiring a large training dataset.

Neither shallow nor deep architectures are suitable for the specific prediction of individual behaviors with a small training dataset; hence, integrating their advantages to design a mixed architecture can be valuable. Considering the task of predicting authors' publication productivity on the dblp dataset$\footnote{See https://www.dblp.org.}$, in this study, we proposed a mixed architecture.
The annual number of publications within a successive time interval can be expressed by a time series taken in that interval. An individual's publication productivity in an interval often correlates with that in long or short time intervals. That is, publication behavior has a long short-term memory. Therefore, we applied a long short-term memory~(LSTM) as a deep architecture to this task\cite{lstm1997}.
For authors without a publication history, we chose a shallow architecture, the piecewise Poisson model, the effectiveness of which has been verified on the dblp dataset for the number of publications for groups of authors on average\cite{Nelder-Wedderburn1972,ref55,ref56}.
Our contribution is the proposal of a model to integrate the advantage of the LSTM with that of the piecewise Poisson model.
The experiments conducted herein show that this mixed architecture integrates the advantages of both shallow and deep architectures and can simultaneously provide fine short-term predictions for individuals and long-term predictions for groups of authors.

The remainder of this paper is organized as follows. A review of the literature and the motivation of this study are described in Sections 2 and 3, respectively. The model and empirical data are described in Sections 4 and 5, respectively. The results and comparisons with previous results are discussed in Sections 6 and 7, respectively. Finally, the conclusions are drawn in Section 8.

\section{Literature review}

The number of citations received by papers or authors, the $h$-index, and publication productivity are the three main aspects for predicting scientific impact\cite{ref55}.
The number of received citations represents the number of times that an academic publication has been cited by other publications\cite{Garfield1955, Harnad2009}.
The $h$-index is a popular measure of scientific success, defined as the maximum value of $h$ such that the given author has published $h$ papers that have each been cited at least $h$ times\cite{Hirsch2005}.
The publication productivity of an author simply represents the number of publications\cite{Way-Morgan2019,ref55}.
However, whereas predicting the citation index has attracted considerable attention, the $h$-index has attracted little attention, and publication productivity has attracted even less attention.

A multitude of methods for predicting these indexes apply mathematical models or specific formulae.
We refer to these methods as shallow architectures, most of which are regression models.
In contrast, deep architectures construct deep networks comprising multiple layers used to progressively extract higher-level features from raw data; these networks have also been applied to predict the above three indexes.
Here, we classified these methods into five categories.

Category $1$: Predicting the number of citations using citation counts directly and indirectly.
Citation-based models consider only the number of citations a paper received within a specific time interval.
Mazloumian predicted highly cited publications based on short-term citations by a multilevel regression model\cite{Mazloumian2012}.
Wang et al derived a mechanistic model for the citation dynamics of individual papers\cite{Wang2013}.
Cao et al applied a Gaussian mixture model to predict the trends of papers' future citations\cite{Cao2016}.
These studies utilized citation counts directly, while some studies have proposed new indexes to indirectly predict the number of citations.
Newman showed that papers with high $z$-scores have the potential to receive more attention than other papers with the same citation counts \cite{Newman2014}.
Pobiedina et al used the temporal evolution of authors' citation networks to define a score for predicting publications' future citations\cite{Pobiedina2016}.
Klimek et al proposed a document centrality indicator\cite{Klimek2016}.

Category $2$: Predicting the number of citations using other factors.
Other factors include journal impact factors, the number of authors, the number of cited references, and the number of pages.
Stern and Abramo utilized linear regression models considering journal impact factors\cite{Stern2014,Abramo2019}.
Kosteas examined the relative strength of short-term citation counts, journal impact factors, and journal rankings in terms of predicting long-run citations\cite{Kosteas2018}.
Bornmann et al calculated the percentiles of citations as an indicator of citation impact in consideration of journal impact factors, the number of authors, the number of cited references and the number of pages\cite{Bornmann2014}.
Hu et al utilized binary classification models with five keyword popularity features to predict whether papers would be highly cited by employing supervised learning\cite{HuTai2020}.
Bai et al introduced a paper potential index model and a multifeature model by analyzing the inherent quality of scholarly papers, scholarly paper impact, early citations, and early citers' impact\cite{Bai2019}.
Yu et al proposed a stepwise regression model by synthesizing the specific features of publications, authors and journals\cite{YuYu2014}.

Category $3$: Using deep architectures to predict the number of citations received.
Abrishami et al utilized an RNN that takes the citation count history of a paper in its early years of publication as the input and predicts the future citation count as the output\cite{Abrishami2019}.
Wang et al applied neural network algorithms combining four classic paper factors: the scientific impact of the first author, the scientific impact of the potential leader, the scientific impact of the team, and the relevance of the authors' existing papers\cite{WangFan2019}.
Ruan et al used a four-layer backpropagation neural network model to predict the five-year citations of papers\cite{RuanZhu2020}.
Xu et al used a CNN to capture the complex nonlinear relationships between early network features and the final cumulative citation count\cite{XuLi2019}.
Mistele et al utilized a feedforward neural network to predict individual authors' future citation counts\cite{Mistele2019}.

Category $4$: Using shallow architectures to predict the $h$-index.
The $h$-index has attracted substantial attention with regard to its prediction.
Ye et al applied a power law model for a specific type of $h$-index sequence by using nonlinear regression\cite{Ye2008}.
Egghe et al showed an existence theorem for the $h$-index and proved that it is a function of the total number of sources\cite{Egghe2006}.
Acuna et al presented formulae to predict the $h$-index and indicated that the current $h$-index is the most significant predictor of its future value compared with the number of current papers\cite{Acuna2012}.
Dong et al used linear regression to predict the future $h$-index; their prediction results show the extent to which authors' future $h$-indexes can be inferred from their previous publication records\cite{Dong2016}.

Category $5$: Predicting publication productivity.
Laurance et al employed a generalized linear model format and found that the number of publications at the completion of the PhD is correlated with long-term publication success\cite{Laurance2013}.
Lehman found the publication productivity tendency to be a curvilinear function of age\cite{Lehman2017}.
Simonton proposed the following prediction formula: $p(t)=c(e^{-at} - e^{-bt})$, where $c=abm/(b-a)$, $a$ is called the ``ideation rate'', $b$ is called the ``elaboration rate'', and $m$ represents the limit number of publications produced by an author in his or her lifetime\cite{Simonton1984}.
Way et al used a piecewise linear model to track productivity trajectories and showed that departmental prestige predicts overall individual productivity\cite{Way-Morgan2017}.
Xie found that the number of publications within a short time interval follows a Poisson distribution when considering authors possessing the same number of publications and then proposed a piecewise Poisson model, which was used to characterize the productivity patterns of many authors\cite{ref55}.

In summary, the success of shallow architectures in predicting the number of citations and the $h$-index results from the cumulative advantage of receiving citations\cite{Price1965}; that is, the future number of received citations positively correlates to the current number\cite{ref55}.
However, the cumulative advantage of publication productivity is weaker than that of the received number of citations; consequently, predicting publication productivity for an individual cannot be accomplished by employing only a shallow architecture regression model\cite{ref55}.

\section{Motivation}

\subsection{Why use the LSTM?}

The annual cumulative number of publications within a successive time interval can be expressed by a time series.
Due to the long- and short-term memory of authors' publication behavior, the publication counts in such time series are correlated with each other.
For long-term memory, for example, it may take a long time to solve a difficult problem; as a consequence, publication productivity from a long ago can influence current publication productivity.
For short-term memory, a research problem focusing on popular topics would motivate the author to produce publications in a short time, as would a new method with good performance on exclusive tasks.
Hence, the current publication productivity of an individual not only is correlated with that in the short preceding time interval but also may be influenced by that of many years prior.

Simple RNNs compose a class of artificial neural networks where connections between nodes form a directed graph along a temporal sequence, which allows the network to exhibit temporal dynamic behavior.
However, when a simple RNN is trained by using backpropagation, the gradients that are backpropagated can ``vanish".
Thus, a simple RNN cannot keep track of arbitrary long-term dependencies in the input sequences.

Among the existing RNN architectures, an LSTM partially solves the vanishing gradient problem.
Additionally, an LSTM is well suited to making predictions based on time series data, especially for time series that are correlated with each other, which is attributable to the cell memory unit in the network.
Specifically, a common LSTM unit is composed of a cell, an input gate, an output gate and a forget gate. The cell remembers values over arbitrary time intervals, while the three gates regulate the flow of information into and out of the cell.

\subsection{ Why use the piecewise Poisson model?}

In a population assumed to follow a Poisson distribution, the probability of the occurrence of an event changes with previous occurrences\cite{Consul1973}.
Previous studies have investigated the distribution of the number of authors' publications, which can be thought of as a mixture of Poisson distributions or a generalized Poisson distribution\cite{XieOu2016,XieOu2018}.
Such a distribution can also be characterized by a trichotomy, including a generalized Poisson head\cite{Xie2019a}.
The event of producing a publication can be regarded as an analogy of observing ``heads'', where the probability of publishing is affected not only by recent events but also by older events\cite{ref55}.
That is, the behavior of producing publications exhibits both long- and short-term memory.

The memory of authors' publication behavior is considered a cumulative advantage.
To weaken the effects of this cumulative advantage, a previous study partitioned authors into several populations to study their publication productivity\cite{ref55,ref56}.
It was found that the number of publications produced by an author within the following short time interval follows a Poisson distribution, which supports the piecewise Poisson model.
Herein, this model was applied to the dblp dataset and exhibited good performance in predicting publication productivity.

\subsection{What are their disadvantages?}

We conducted an experiment to examine the performance of the LSTM; the result shows that the LSTM is not capable of predicting the publication productivity of authors~(Appendix~A).
This failure can be explained as a result of the feature of deep architectures that the neural network regards all authors who produced the same numbers of publications in his or her history as equally productive in the future. For new authors rising in academia, LSTM would indicate that they less productive since there are no publications in their history.
Thus, applying LSTM to directly predict publication productivity for all authors appears undesirable.

In contrast, the piecewise Poisson model is applicable for a group of authors on average, and precisely predicts publication events for low-productivity authors.
However, it is not suitable for predicting individuals' publication productivity in long-term intervals and does not precisely predict publication events for high-productivity authors\cite{ref55}.
This model is inapplicable for individual authors both because an individual's research is random and because of the nature of regression.

In summary, the LSTM has both long- and short-term memory and is suitable for making predictions based on the time series of the annual number of publications; however, it cannot predict the number of authors' publications directly.
The piecewise Poisson model can determine the publication productivity patterns of many authors and exhibits good performance in long-term prediction for groups of authors; however, for individual authors, the model does not perform well on long-term prediction.
That is, applying a deep or shallow architecture alone cannot achieve good performance on the task of predicting individuals' publication productivity.
Therefore, we proposed a model integrating both architectures and conducted experiments to examine its performance on this prediction task.

\section{The model}

\begin{algorithm}
\caption{Predicting the number of publications.}
\label{algmodel}
\begin{algorithmic}

\REQUIRE ~~\\
the time series \{$h_s(t_{X-12})$, $h_s(t_{X-11})$, ..., $h_s(t_{X-1})$\} of any tested author $s$;\\
parameters $q$, $\beta_1$, $\beta_2$.\\

\ENSURE ~~\\
the predicted number of publications $\hat{h}_s (t_Y)$ of author $s$.

\FOR{each author $s$}
\STATE{input the time series into the LSTM to predict $\overline h_s(t_X)$;}
\STATE{sample $x$ from Power-law($h_s(t_{X-1})$, $q$, $\beta_1$, $\beta_2$);}
\STATE{sample $y$ from Pois$(x)$;}
\STATE{let  $\hat{h}_s(t_X)$= $\overline h_s(t_X)$ + $y$;}
\FOR{$l$ from $X+1$ to $Y$}
\STATE{sample $x$ from Power-law($\hat{h}_s(t_{l-1})$, $q$, $\beta_1$, $\beta_2$);}
\STATE{sample $y$ from Pois$(x)$; }
\STATE{let  $\hat{h}_s(t_l)$= $\hat{h}_s(t_{l-1})$ + $y$;}
\ENDFOR

\ENDFOR

\end{algorithmic}
 \end{algorithm}

  \begin{figure*}[ht]
\centering
\includegraphics[height=3.0     in,width=6.0    in,angle=0]{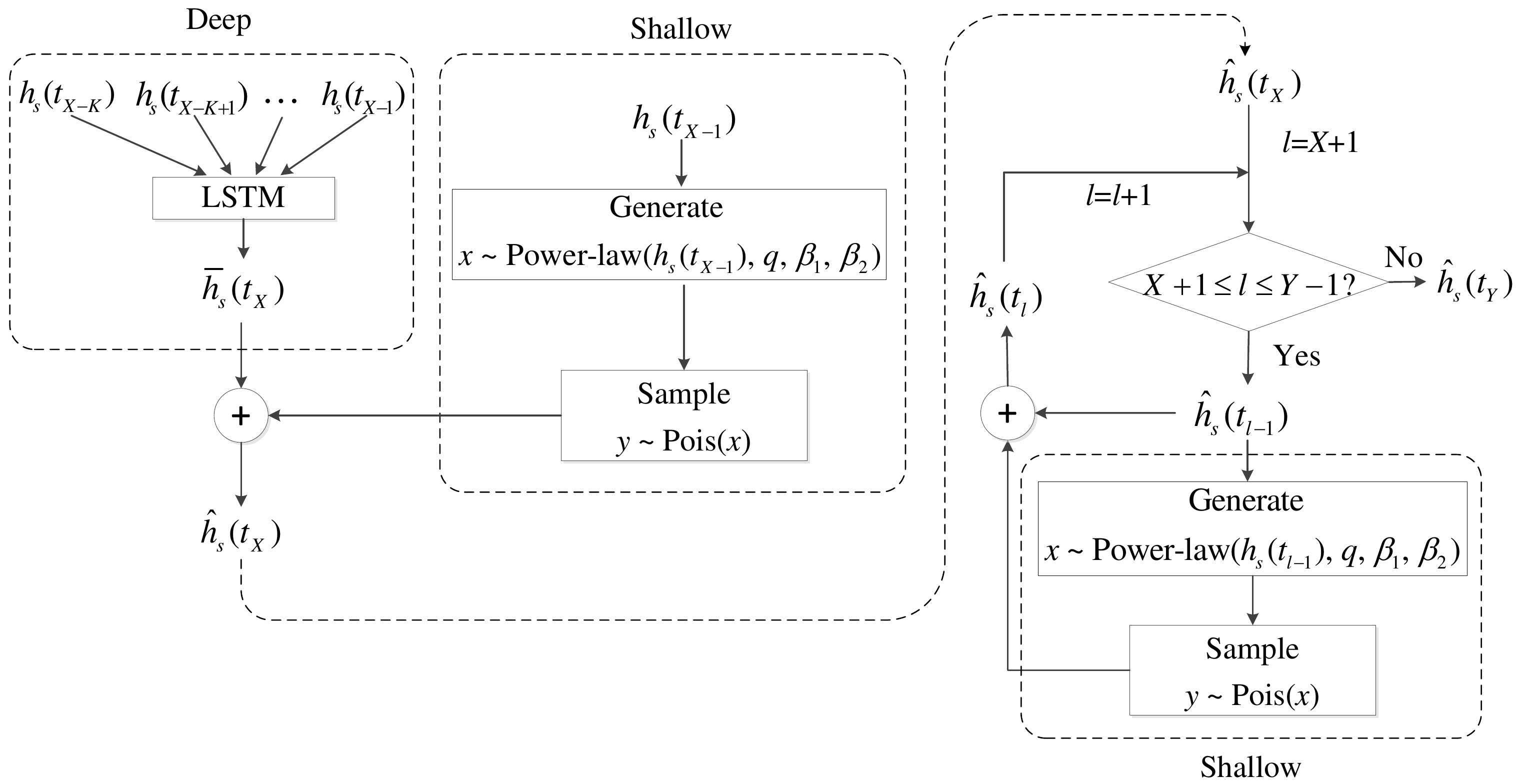}
\caption{   {\bf  The flowchart of the proposed model.}
{The steps and the intermediate input/output of Algorithm \ref{algmodel} are illustrated. The left framework is the LSTM, and the right two frameworks are shallow architectures.
A plus sign in a circle expresses the addition of  two inputs. The details regarding each step are described in the ``The model'' section.
}
}
 \label{flow}
 \end{figure*}

We proposed a model integrating the advantage of the LSTM and that of the piecewise Poisson model~(see Algorithm~\ref{algmodel}).
Here, we illustrated model terms, described the deep architecture and the shallow architecture, and provided the details of integrating the two architectures.

\subsection{Model terms}

Consider authors with their numbers of publications in years $T_0$--$T_1$ and $T_1$--$T_2$.
The number of publications produced by author $s$ in years $T_0$--$t_l$ is denoted $h_s(t_l)$, where $T_0<t_l<T_2$.
That is, $h_s(t_l)$ represents the cumulative number of publications produced by author $s$ from year $T_0$ to $t_l$.
The time series \{$h_s(t_{X-12})$, $h_s(t_{X-11})$, ..., $h_s(t_{X-1})$\} is constructed as the input of our model.
$\overline h_s(t_{X})$ represents a temporary prediction for $h_s(t_{X})$.
$\hat{h}_s(t_X)$ represents the sum of $\overline h_s(t_{X})$ and $y$, and $\hat{h}_s(t_l)$ represents that of $\overline h_s(t_{l})$ and $y$, where $X+1 \le l \le Y$.
The aim of our model is to predict $\hat {h}_s (t_Y)$, the number of publications produced by authors $s$ in years $T_0$--$t_Y$.

\subsection{Deep architecture: the LSTM}

We used the ``Keras'' framework to design an LSTM network, which possesses $32$ LSTM units.
Given a time series of $12$ numerical data points, the LSTM outputs one value.
A ``Dense" layer is applied to generate one output.
Rectified Linear Units~(ReLU)\cite{ReLU2010}, a nonlinear activation function, is used to calculate the activation values.
Root Mean Square prop~(RMSprop), an adaptive learning rate method, is chosen to optimize the LSTM network.
Mean Square Error~(MSE), the average squared difference between the estimated values and the actual value, is used as the loss function.
The network has 4,385 parameters calculated by the fourfold cross validation with the time series \{$h_s(t_{X-12})$, $h_s(t_{X-11})$, ..., $h_s(t_{X-1})$\} and the target $h_s(t_{X})$ of any author $s$ in the training dataset.
For fourfold cross validation, the training dataset is randomly divided into 4 packets: one of the packets is used as the test set, while the remaining 3 packets are used as the training set. The training/validation accuracy is the average of that of the 4 models.
The data are fed to the LSTM in batches of 5 authors~(batch-size=5).

\subsection{Shallow architecture: the piecewise Poisson model}

Suppose that a random variable $z$ follows a power-law distribution of parameters $h$, $q$, $\beta_1$ and $\beta_2$.
The power-law distribution can be expressed as Power-law($h$, $q$, $\beta_1$, $\beta_2$) in Algorithm~\ref{algmodel}, where $h$ is the output of the LSTM model and the remaining three parameters are constants.
The probability density function of the power-law distribution is as follows,
\begin{equation}
{f(z, h, q, \beta_1, \beta_2)=\dfrac{q}{(\beta_1 h^{\beta_2})^q} z^{q-1}, }
\label{powerlaw}
\end{equation}
where $0 \le z \le \beta_1 h^{\beta_2}$.
We sample $x$ from the power-law distribution.
Notably, we used $\beta_1 h^\beta_2$ as the scale parameter of the power-law distribution, the aim of which is to express the heterogeneity of authors' publication ability.

Then, we introduced a random variable $w$, which follows a Poisson distribution with $x$ as the expectation.
The probability function of $w$ is given by
\begin{equation}
{p(w=k)=\dfrac{x^k}{k!} \mathrm e^{-x}, }
\label{poisson}
\end{equation}
where $k=0, 1, 2, ...$.
The Poisson distribution can be expressed as Pois($x$) in Algorithm~\ref{algmodel}.
We sampled $y$ from the Poisson distribution with $x$ as its expectation, which is based on the fact that the number of publications produced by an author within the following short time interval follows a Poisson distribution\cite{ref55}.

\subsection{Implementation details}

We summarized the details of our model to clarify how we integrate the deep architecture with the shallow architecture.
Algorithm~\ref{algmodel} is provided to predict the future cumulative number of publications of authors in the time interval $[t_X, t_Y]$, where $T_1 \le t_X < t_Y \le T_2$.
To make our model easily accessible, we provided a generalized flowchart corresponding to Algorithm~\ref{algmodel}, as illustrated in Fig.~\ref{flow}.
In the LSTM, $K$ denotes the length of the time series, and here, we let $K$ be 12.
The implementation details are described below.

For the first year $t_X$, with the time series \{$h_s(t_{X-12})$, $h_s(t_{X-11})$, ..., $h_s(t_{X-1})$\} as its input, the LSTM output $\overline {h}_s(t_X)$.
Meanwhile, we sampled $x$ from the power-law distribution with its parameter $h$ being $h_s(t_{X-1})$.
Then, we sampled $y$ from the Poisson distribution with $x$ as its expectation.
Having obtained $\overline {h}_s(t_X)$ and $y$, we took their sum as the predicted cumulative number of publications, namely, $\hat{h}_s(t_X)$.
Given $\hat{h}_s(t_{X})$, we continued to predict the cumulative number of publications for the following years.
For the year $t_{l}$, where $X+1 \le l \le Y$, we only utilized the piecewise Poisson model.
We sampled $x$ from the power-law distribution with its parameter $h$ being $\hat{h}_s(t_{l-1})$.
Then, we sampled $y$ from the Poisson distribution with $x$ as its expectation.
Similarly, we took the sum of $\hat{h}_s(t_{l-1})$ and $y$ as the predicted value of the cumulative number of publications in the current year.
After several iterations, we obtained $\hat{h}_s(t_{Y})$.

\section{Empirical data}\label{data}

The dblp computer science bibliographic dataset was applied in this study, which provides open bibliographic information on most  of the journals and conference proceedings in computer science.
The quality of this dataset is guaranteed by a range of measures,  such as applying  several  methods of name disambiguation,
linking 60,000 manually confirmed external IDs to dblp author bibliographies, and so on.
The dataset is comprised of 315,677 publications produced by 441,501 authors, which have been published in 1,701 journals and conference proceedings in the time interval 1951--2019.

We extracted parts from the dataset dblp to construct training and test datasets for the experiments in Section 6.
The training dataset consists of 5,741 authors who produced publications at year 2000 and their annual number of publications in years 1951--2013,  involving 105,806 publications.
The test dataset consists of those authors and their annual number of publications in years 2014--2018, involving 111,612 publications.

Following data processing, we obtained the time series of the number of publications by each author, in which each datum represents the number of an author's publications produced in the years from 1951 to the current year.
In details, here the $h_s(t)$ in the model is the number of publications produced by author $s$ in years 1951--$t$.
The proposed model is trained on every author $s$ in the training dataset, where the time series \{$h_s(2001)$, $h_s(2002)$, ..., $h_s(2012)$\} is the input and $h_s(2013)$ is the target.
The time series input for testing our model is \{$h_s(1989)$, $h_s(1990)$, ..., $h_s(2000)$\}.
Notably, the annual number of papers in years 1951--2013 has been used in training process, the annual number  used for test purpose is only that at years 2014--2018.

\section{Results}\label{Results}

 \begin{figure*}[ht]
\centering
\includegraphics[height=3.25     in,width=6    in,angle=0]{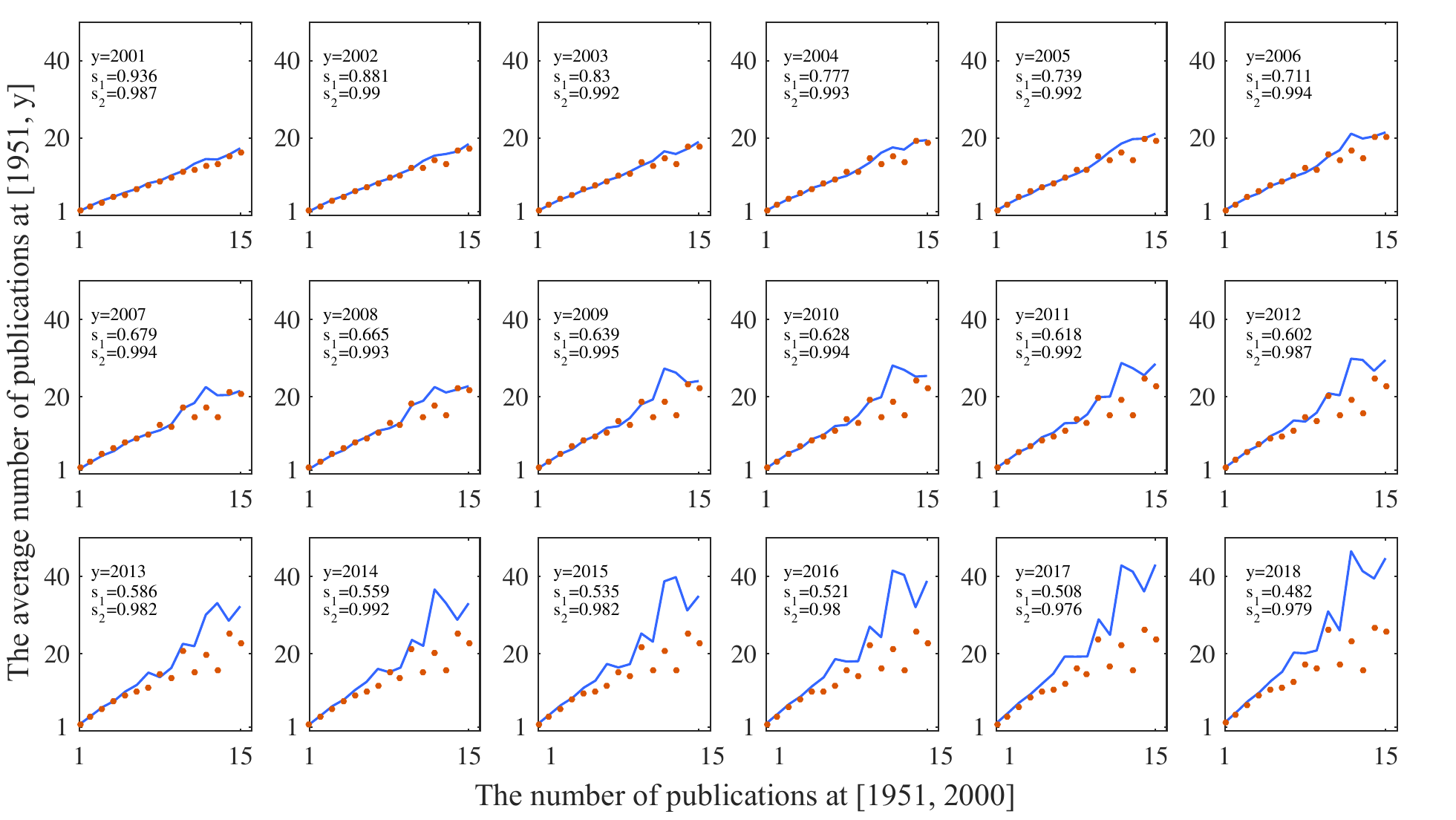}
\caption{   {\bf  The trend   of  the number of   publications predicted by the proposed model.}
The tested authors are those who have $i$ publications within $[1951,2000]$, where $i=1,...,15$.
Red dots represent the average numbers of their publications within $[1951,y]$ ($n(i,y)$), and blue lines represent the predicted numbers ($m(i,y)$).
Index $s_1$ is the Pearson correlation coefficient, which is calculated based on the list of the number of publications by an author and the list of their predicted numbers. Index $s_2$ is this coefficient based on the sorted lists.}
 \label{model-trend-2000}
 \end{figure*}

  \begin{figure*}[ht]
\centering
\includegraphics[height=3.2     in,width=6.2     in,angle=0]{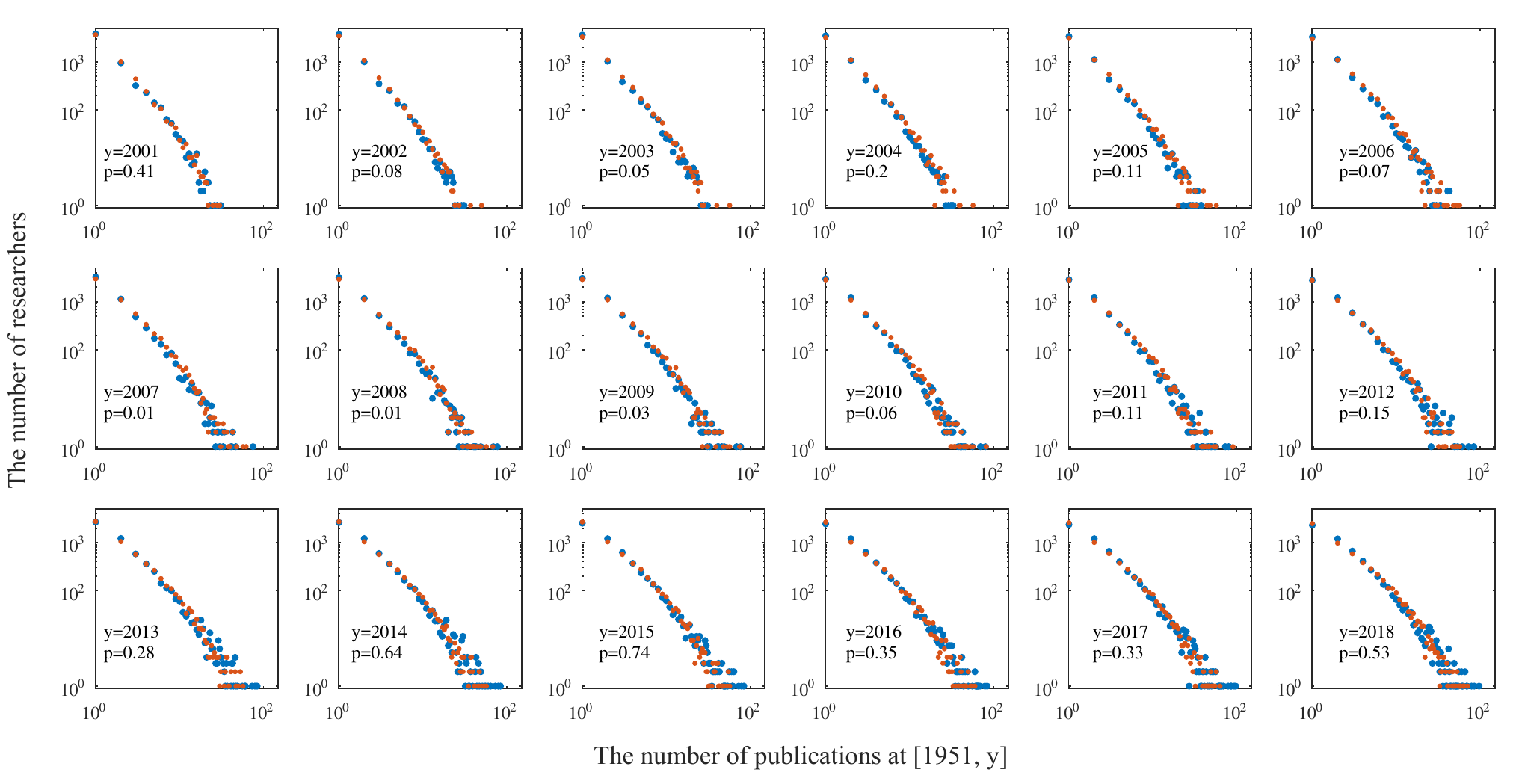}
\caption{   {\bf  The   distribution of the number of publications predicted by the proposed model.} Red circles represent the ground truth for the tested authors, and blue circles represent the predicted distributions. The $p$-value is from the KS test with the null hypothesis that the compared distributions are the same.
}
 \label{model-distr-2000}
 \end{figure*}

  \begin{figure*}[ht]
\centering
\includegraphics[height=3.35     in,width=6.2     in,angle=0]{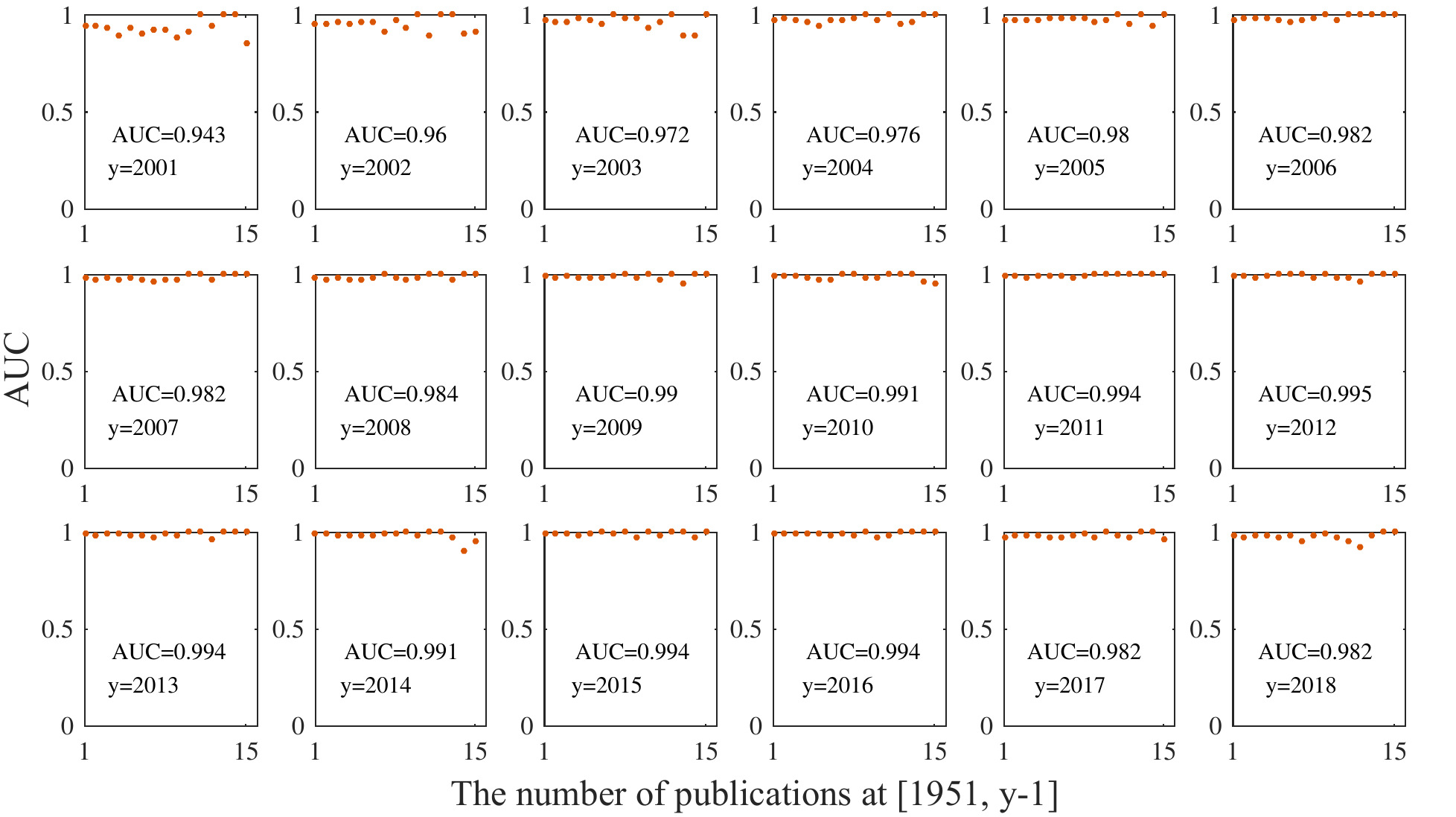}
 \caption{     {\bf  The short-term precision of predicting the probability of publication.} Red dots represent the AUC of predicting the publication events in year $y$ for the tested authors who have $i$ publications within $[1951, y-1]$, where $i=1,...,15$. The inputs are the true numbers of publications.
The AUC was calculated based on all of the tested
authors.         }
 \label{model-auc-annual-2000}
 \end{figure*}

 \begin{figure*}[ht]
\centering
\includegraphics[height=3.2     in,width=6.2     in,angle=0]{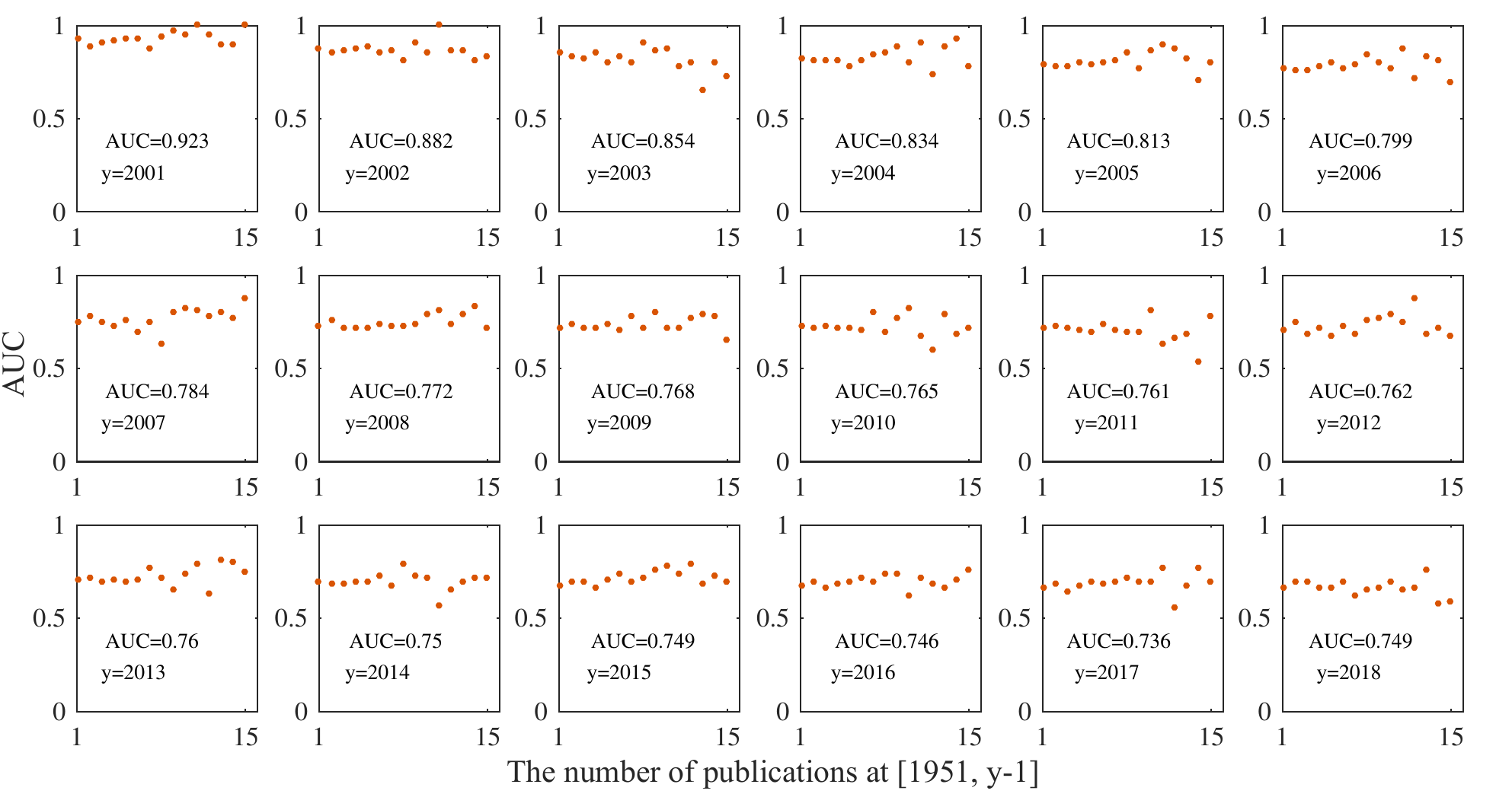}
 \caption{ {\bf  The long-term precision of predicting the probability of publication.} Red dots represent the AUC of predicting the publication events in year $y$ for the tested authors who produced $i$ publications within $[1951, y-1]$ predicted by our model, where $i=1,...,15$, $y=2002, 2003, ..., 2018$.
For the tested authors who produced $i$ publications within $[1951,2000]$, the input is the true number of publications.
The AUC was calculated based on all of the tested
authors.         }
 \label{model-longterm-auc-2000}
 \end{figure*}

We tested the performance of our model in predicting publication productivity in three ways, namely, the evolutionary trend of the number of publications, the distribution of the number of publications, and the probability of producing publications.
Since predicting for extremely productive authors is beyond the ability of our model, we only considered the authors who produced no more than 15 papers in years 1951--2000, which count for about 94\% authors in the dblp dataset.
Notably, the hyperparameters in the shallow architecture are $q=0.1$, $\beta_1=0.33$, and $\beta_2=1.22$.

\subsection{Predicting the trend of producing publications}

Consider the tested authors who have $i$ publications in years 1951--2000, where $i=1,...,15$.
Denote the average number of publications produced by those authors in years 1951--$y$ as $n(i,y)$.
Denote the predicted average number of publications as $m(i,y)$.
Fig.~\ref{model-trend-2000} shows the trend of the predicted average number of publications and that of the ground truth.
Given year $y$, these two trends with respect to $i$ can be seen.
We measured the correlation between these two trends by applying the Pearson correlation coefficient\cite{Hollander} on the individual level ($s_1$) and on the group level ($s_2$).
The former is calculated based on the list of the numbers of publications and on that of the predicted numbers, while the latter is based on the sorted lists.
Index $s_1$ exhibits a gradual decreasing trend, while index $s_2$ remains high.
This result shows that our model is applicable to short- and long-term predictions for a group of authors on average, but it cannot deal with long-term predictions for individuals.

\subsection{Predicting the distribution of the number of publications}

The distribution of the predicted numbers of publications can be compared with that of the real numbers to inspect the capability of our prediction model at the group level.
Consider the tested authors and their publications produced in years 1951--$y$, where $y=2001,...,2018$.
Fig.~\ref{model-distr-2000} shows the evolution of the true numbers of publications and that of the predicted numbers of publications.
There appears to be a fat tail in the real number of publications, which can be explained by a small group of extremely productive authors.
The distribution predicted by our model also exhibits a fat tail that can fit the ground truth well.
We used the Kolmogorov-Smirnov~(KS) test to compare the two observed distributions.
Here, the null hypothesis is that the two distributions are identical, and the threshold value for determining statistical significance is 0.05.
As the $p$-value remains above 0.05, the KS test cannot reject the null hypothesis, either in their foreparts or in the long-term future.
This finding indicates that our model is able to predict productive authors' publication productivity.

\subsection{Predicting the probability of publication in the following year}

We also tested the performance of our model in predicting the probability of authors producing publications in the following year.
We measured the short-term prediction precision, which is different from the above two experiments.
The area under the receiver operating characteristic curve (AUC) is used to measure the prediction precision\cite{ref55,ref56}.

If an author did (did not) produce publications in the following year, the probability was defined as being greater (less) than $0.5$.
Denote the counts of occurrences of the above two situations (when authors did and did not produce publications in the following year) as $m_1$ and $m_2$, respectively, and denote the count of the times that the probability is $0.5$ as $m_3$.
Denote the number of tested authors as $m$.
The AUC is defined below:
\begin{equation}\mathrm{AUC}= \frac{m_1+m_2+0.5m_3}{m}.\label{auc}\end{equation}
Figs.~\ref{model-auc-annual-2000}-\ref{model-longterm-auc-2000} show the AUC values calculated by formula~(\ref{auc}).

Consider the tested authors who produced $i$ publications in years 1951--$y-1$, where $i=1,...,15$ and $y=2001,...,2018$.
The AUC in Fig.~\ref{model-auc-annual-2000} represents the short-term precision of predicting the probability of authors producing publications in the following year, which indicates that the precision is higher than that of predicting the number of publications.
As this figure shows, for both authors with a small number of historical publications and those with the majority of publications, the AUC value remains high, which indicates the good performance of our model in predicting publication events the following year for all authors.

Consider the tested authors who produced $i$ publications in years 1951--2000 and those who produced $i$ publications in years 1951--$y-1$ predicted by our model, where $i=1,...,15$ and $y=2002,...,2018$.
The AUC in Fig.~\ref{model-longterm-auc-2000} represents the long-term precision of predicting the probability of authors producing publications next year, showing that the probability decreases over time.

\section{Discussion and conclusions}

We proposed a model to predict the number of publications by authors that integrates the advantage of LSTM with that of the piecewise Poisson model.
Due to its long- and short-term memory, the LSTM was used as the deep architecture in the model with a time series as its input.
However, the LSTM predicts new authors as having no publications in future years.
To ameliorate this problem, the sum of a number sampled from a Poisson distribution and the LSTM output is taken as the predicted number of publications for each author.
The dblp dataset is used to test the practicability of our model.
The prediction results indicate the good performance of our model in predicting the number of publications over a relatively long time interval for a group of authors on average. Our model also performs well in predicting publication events in the following year for individuals.

The proposed model attempts to integrate a deep architecture with a shallow architecture to predict publication productivity.
This integration improves the precision of predicting publication events for highly productive authors.
Compared to the models proposed in previous studies, our model outperforms in the short-term prediction for individual authors, which can be attributed to the long-term memory of the LSTM and the characteristics of many authors' publication patterns provided by the piecewise Poisson model.

The proposed model has potential value in specific situations.
In particular, predicting the publication productivity of authors is a basic task for academic administrators and funding agencies.
The proposed model provides a method that would be valuable for performing this task.
For example, funding agencies could apply this model to evaluate the capability of an author to produce publications in the future.
The model would also be practical for disciplinary assessment agencies to predict the scientific performance of a department.
In addition, the model can provide more insights into understanding the evolution of coauthorship networks because of the influence of publication productivity on collaborations in scholarly communication\cite{Xie2020coauthors}.
At least, the proposed model is suitable for predicting publication productivity for authors in the field of computer science.

Nevertheless, some limitations in our model remain.
First, the hyperparameters in our model are set empirically. Thus, a generalized method is expected to be designed to estimate the hyperparameters based on the given dataset.
This indicates a direction for improving our model.
Second, in the distributions of the numbers of publications, the $p$-values of the KS test tend to decline over time because few authors are highly productive; that is, the majority of authors are less productive.
Third, accurate long-term predictions are difficult to achieve because some factors may directly or indirectly affect the productivity of authors, such as the number of coauthors in past publications and the rank of the institutions to which the authors belong.
In future work, we expect to take related factors into consideration.

\section*{Acknowledgments}

This work is supported by the National Natural Science Foundation of China (Grant No. 61773020) and the National Education Science Foundation of China (Grant No. DIA180383).

\begin{figure*}[ht]
\centering
\includegraphics[height=3.2     in,width=6.2     in,angle=0]{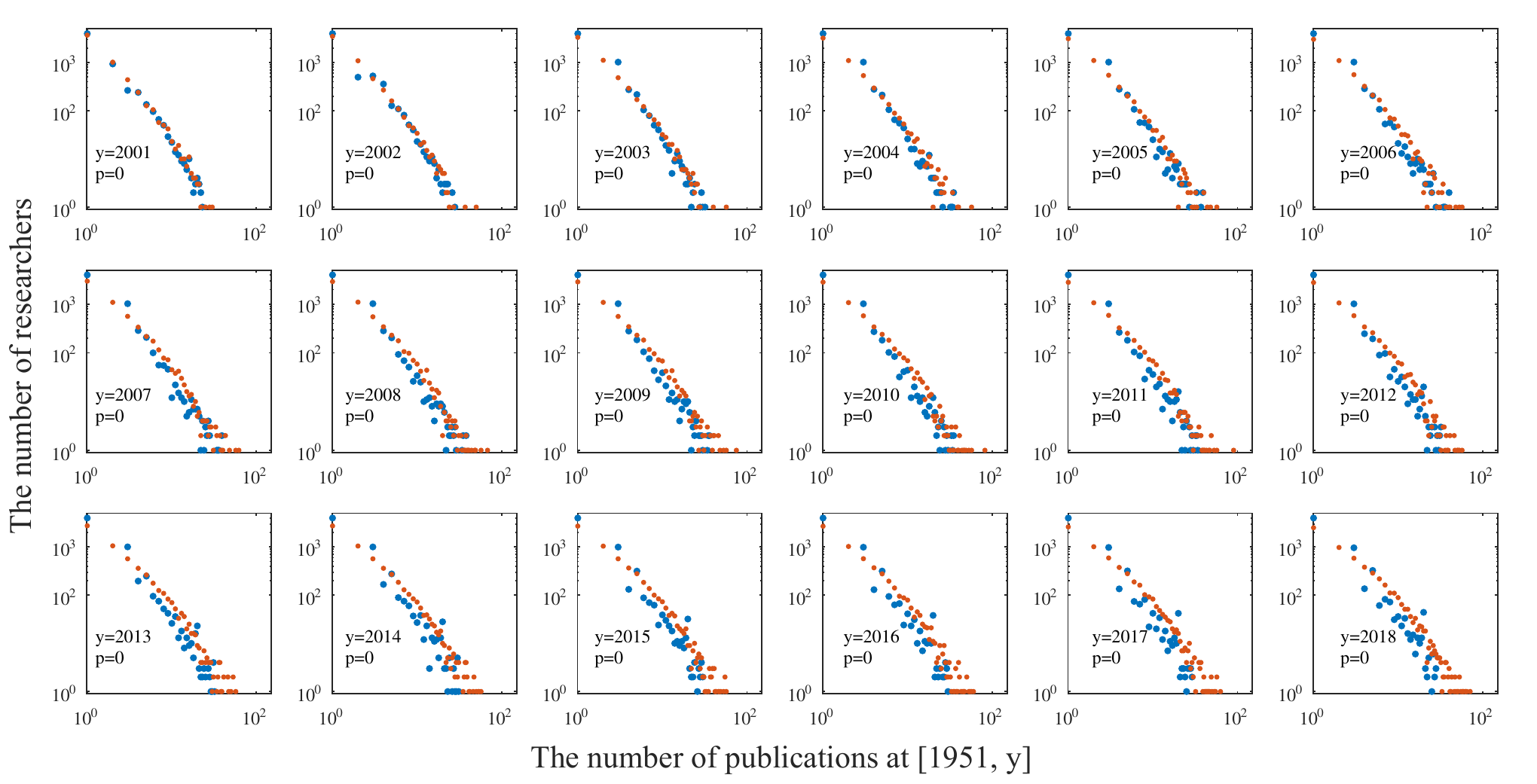}
\caption{   {\bf  The   distributions of the number of publications.} Red circles represent the ground truth for the tested authors, and blue squares represent the distributions predicted by the LSTM. The $p$-value is from the KS test with the null hypothesis that the compared distributions are the same.
}
 \label{lstm-distr-2000}
 \end{figure*}

  \begin{figure*}[ht]
\centering
\includegraphics[height=3.2     in,width=6.2     in,angle=0]{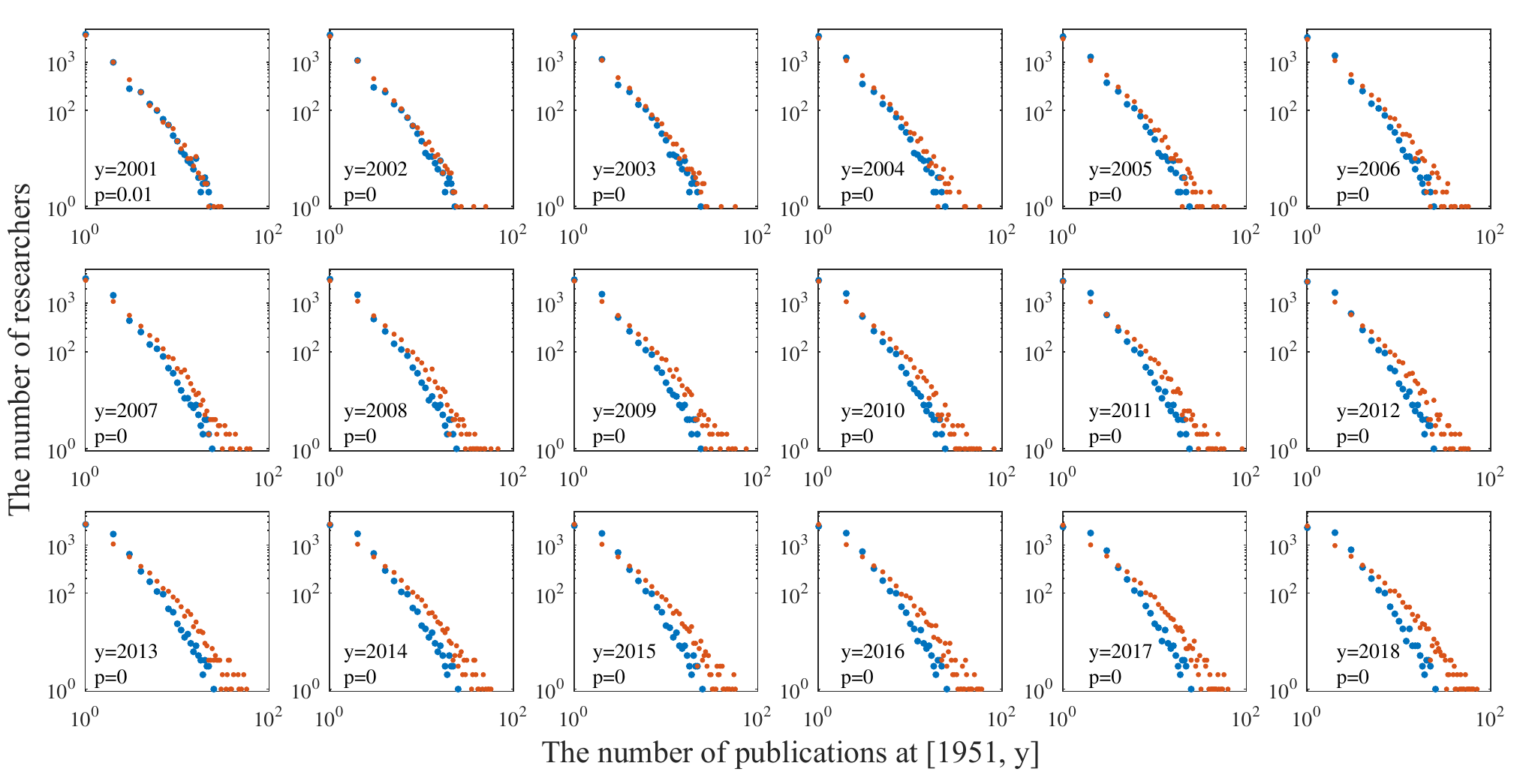}
\caption{   {\bf  The distributions of the number of publications.} Red circles represent the ground truth for the tested authors, and blue squares represent the predicted distributions. The $p$-value is from the KS test with the null hypothesis that the compared distributions are the same.
}
 \label{lstm-poi-distr-2000}
 \end{figure*}

  \begin{figure*}[ht]
\centering
\includegraphics[height=3.2     in,width=6.2     in, angle=0]{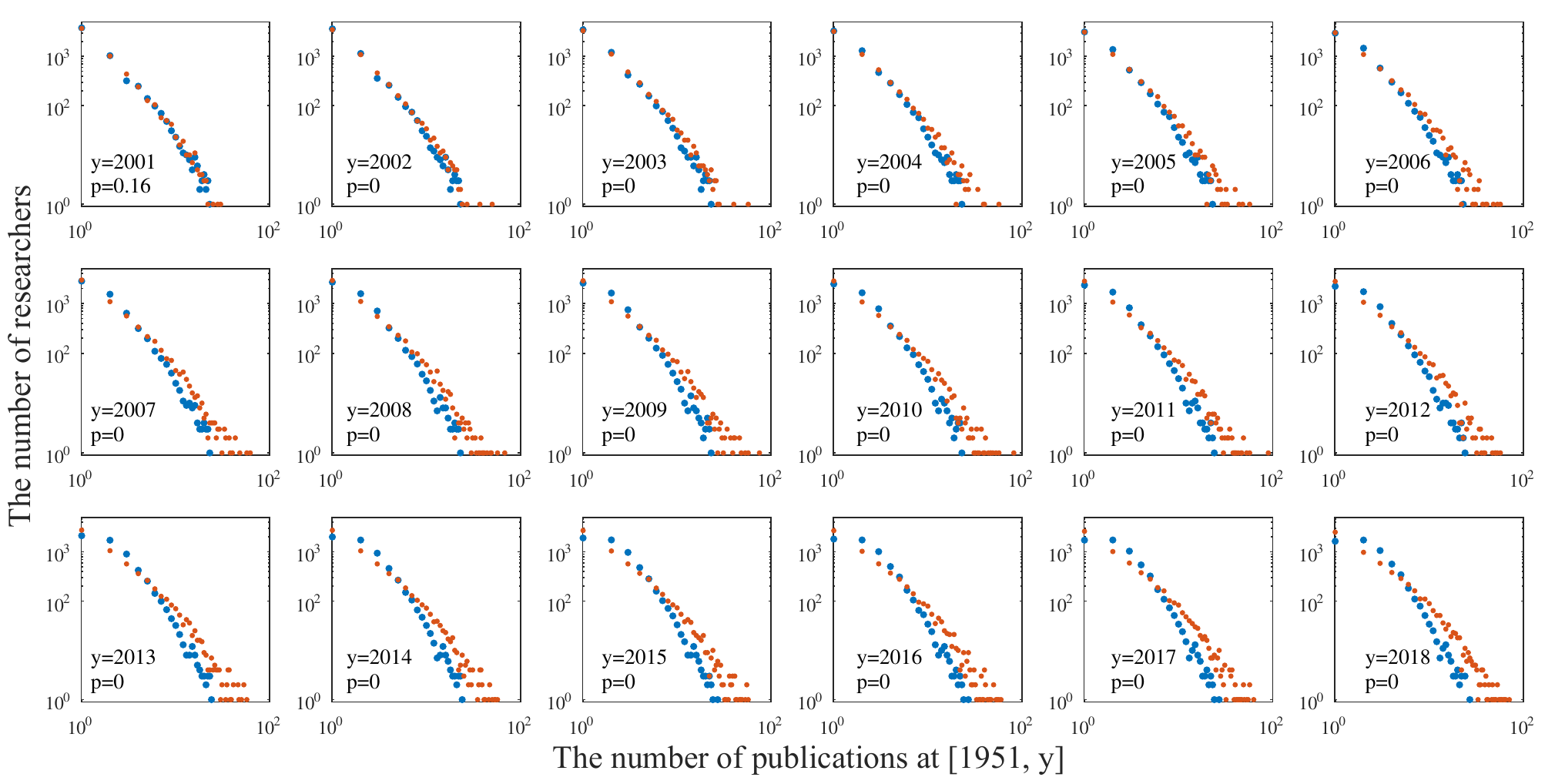}
\caption{   {\bf  The distributions of the numbers of publications.} Red circles represent the ground truth for the tested authors, and blue squares represent the predicted distributions. The $p$-value is from the KS test with the null hypothesis that the compared distributions are the same.
}
 \label{lstm-poi-powerlaw-distr-2000}
 \end{figure*}

\section*{Appendix A: The explanations for the architectures}\label{Appendix A}

We conducted three experiments to predict the number of publications by authors, to explain the role of the architectures in our model.
The first experiment masked the shallow architecture; the second experiment masked the power-law distribution;
and the third experiment masked the scale parameter of the power-law distribution.
The training datasets used in these three experiments are the same as those in our model, as are the test datasets.

\subsection*{Masking the shallow architecture}

We used the deep architecture of our model directly to predict the number of publications.
The deep architecture, the LSTM, was applied to the test dataset, with the time series $\{h_s(1989), h_s(1990), ..., h_s(2000)\}$ as its input.
The result indicates that the LSTM fails to accurately predict publication productivity for authors, and the predicted values lag behind the ground truth~(see the $p$-value in Fig.~\ref{lstm-distr-2000}).
The KS test rejects the null hypothesis that the two distributions are identical.

\subsection*{Masking the power-law distribution}

We sampled a number from a Poisson distribution with its expectation being a constant, of which the probability function is shown in formula~(\ref{poisson}).
For the first year $t_X$, the number sampled from the Poisson model was added to the output of the LSTM, and the result was taken as the predicted number of publications in years $T_0$--$t_X$.
From year $t_l$, where $X+1 \le l \le Y$, we also sampled a number from the Poisson distribution.
Then, we took the sum of the number and the predicted number of publications within $[T_0, t_{l-1}]$ as the predicted number of publications in years $T_0$--$t_l$.
Compared to the result in the first experiment, the fit on the true distribution is significantly improved, but the gap between the two distributions gradually increases over time~(see Fig.~\ref{lstm-poi-distr-2000}).
Specifically, the critical difference between the two distributions is highlighted in their tails.
To generate a fat tail in the predicted distribution, we conducted the third experiment.

\subsection*{Masking the scale parameter of the power-law distribution}

We sampled a number from a power-law distribution with the scale parameter in formula~(\ref{powerlaw}) being 1.
Then, we sampled an integer from the Poisson distribution with its expectation being the number sampled from the power-law distribution.
The remaining steps are the same as those in the second experiment.
The result outperforms that in the second experiment in predicting the distribution of the number of publications~(see Fig.~\ref{lstm-poi-powerlaw-distr-2000}).

\section*{Appendix B: Comparisons with previous work}

We compared our model with three models in previous work in predicting the number of publications by authors.
The three models are the piecewise Poisson model, the model in reference \cite{ref56}, and the GRU.
We conducted the comparisons in three ways, namely, the evolutionary trend of the number of publications, the distribution of the number of publications, and the probability of producing publications.
The results show that the performances of our model, the piecewise Poisson model and the model in reference \cite{ref56} are comparable in long-term prediction.
Our model outperforms the other two models in short-term prediction, which shows the advantage of integrating LSTM with the piecewise Poisson model.
The GRU exhibits poor performance in predicting publication productivity, although the short-term precision in predicting publication events is high.
The principles of the three models and the comparison details are described below.

\subsection*{The piecewise Poisson model}\label{The piecewise Poisson model}

 \begin{figure*}[ht]
\centering
\includegraphics[height=3.05 in,width=6 in,angle=0]{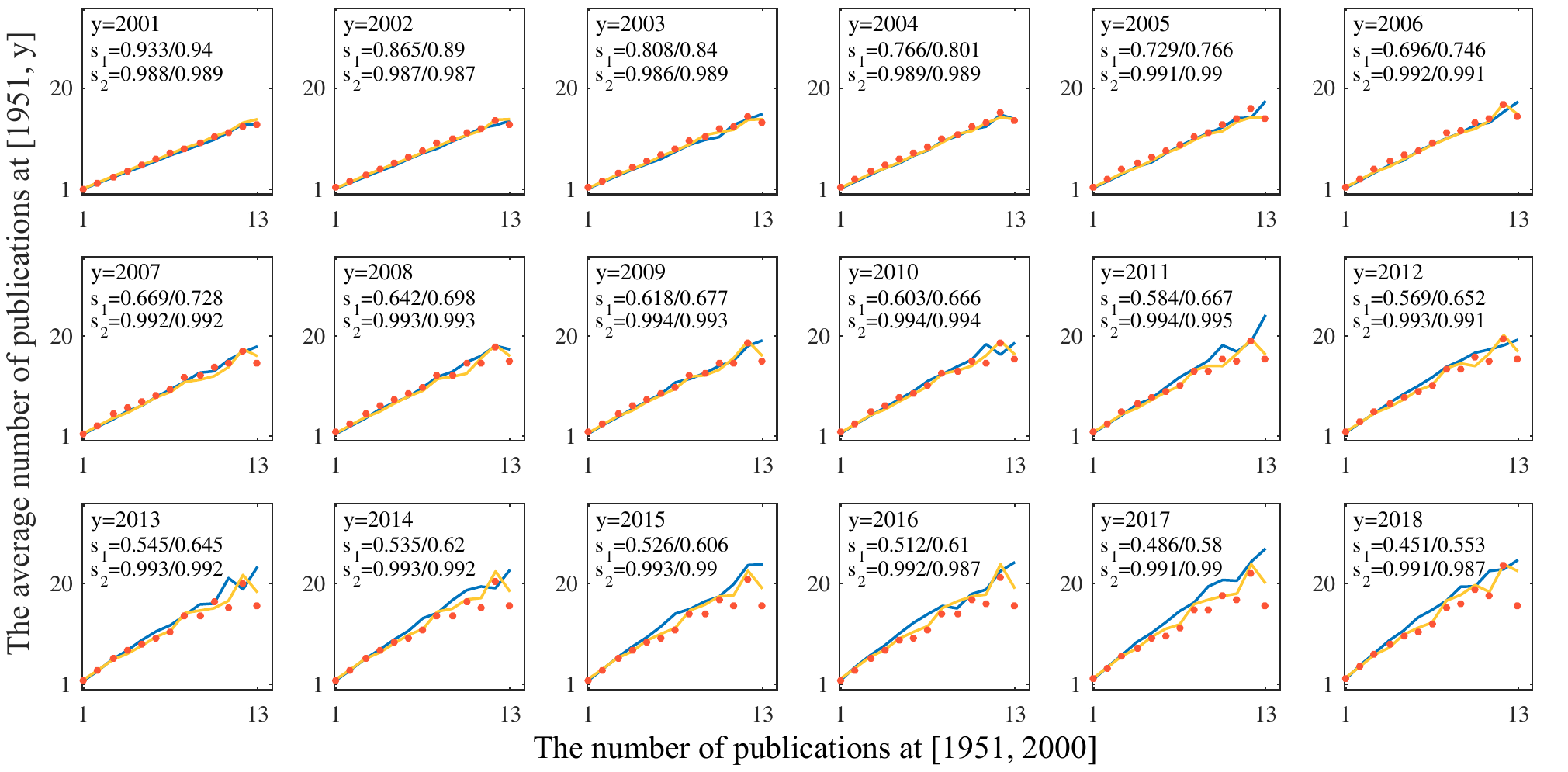}
\caption{   {\bf The trends of the number of publications.}
The tested authors are those who have $i$ publications within $[1951,2000]$, where $i=1,...,13$.
Red dots represent the average numbers of their publications within $[1951,y]$, blue lines represent the numbers predicted by the piecewise Poisson model, and yellow lines represent the numbers predicted by our model.
Indexes $s_1$ and $s_2$ are the Pearson correlation coefficients defined in Fig.~\ref{model-trend-2000}, where the values on the left of ``/" are calculated by our model and the values on the right of ``/" are calculated by the piecewise Poisson model.}
 \label{model_poisson_trend_compare}
 \end{figure*}

  \begin{figure*}[ht]
\centering
\includegraphics[height=3.2     in,width=6.2     in,angle=0]{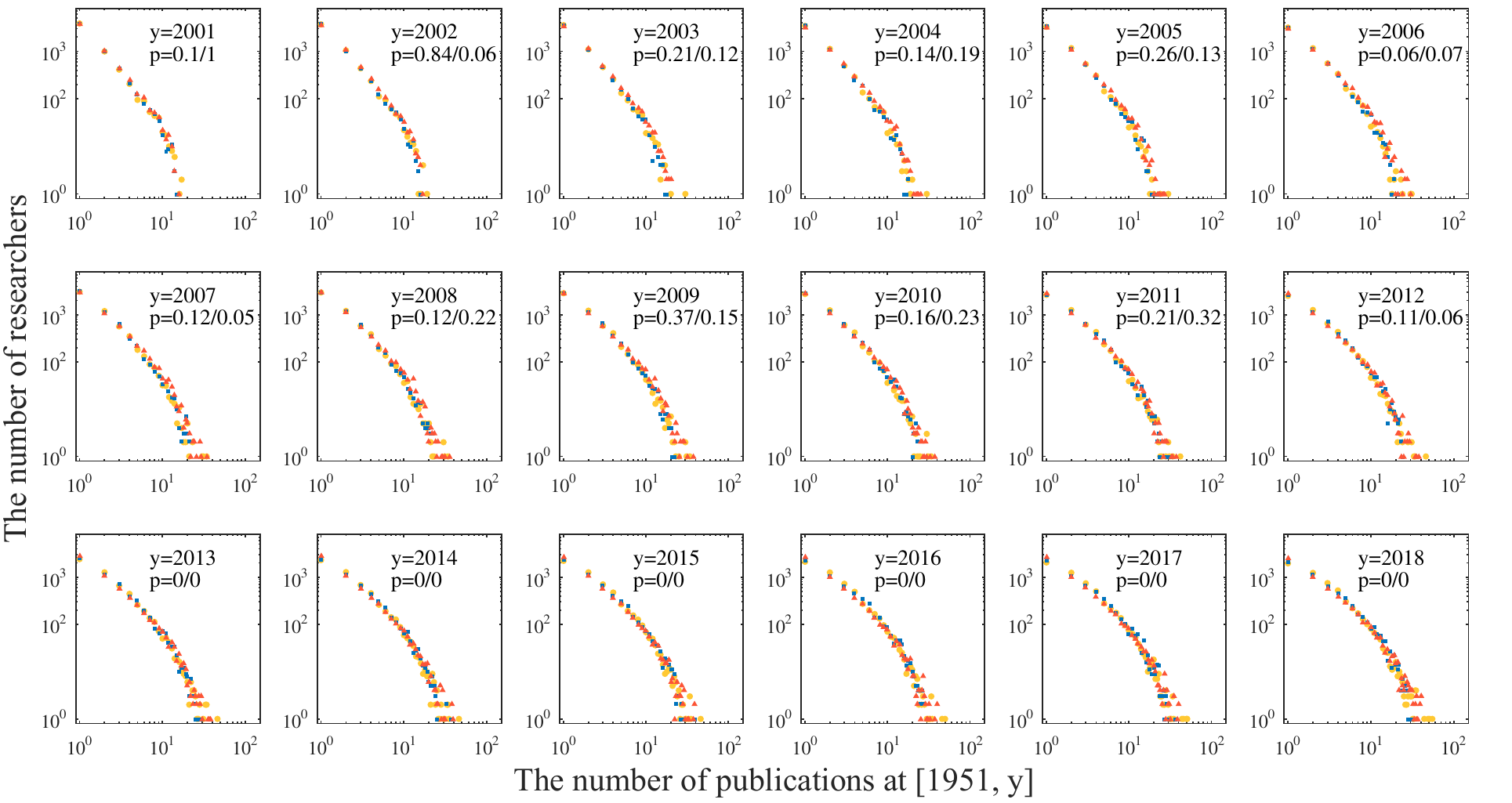}
\caption{   {\bf  The   distribution of the number of publications.} Red triangles represent the ground truth for the tested authors, blue squares represent the distributions predicted by the piecewise Poisson model, and yellow circles represent the distributions predicted by our model. The $p$-value on the left of ``/" is from the KS test with the null hypothesis that the predicted distribution by our model and the true distribution are the same, and that on the right of ``/" is for the piecewise Poisson model.
}
 \label{model_poisson_distribution_compare}
 \end{figure*}

  \begin{figure*}[ht]
\centering
\includegraphics[height=3.35     in,width=6.2     in,angle=0]{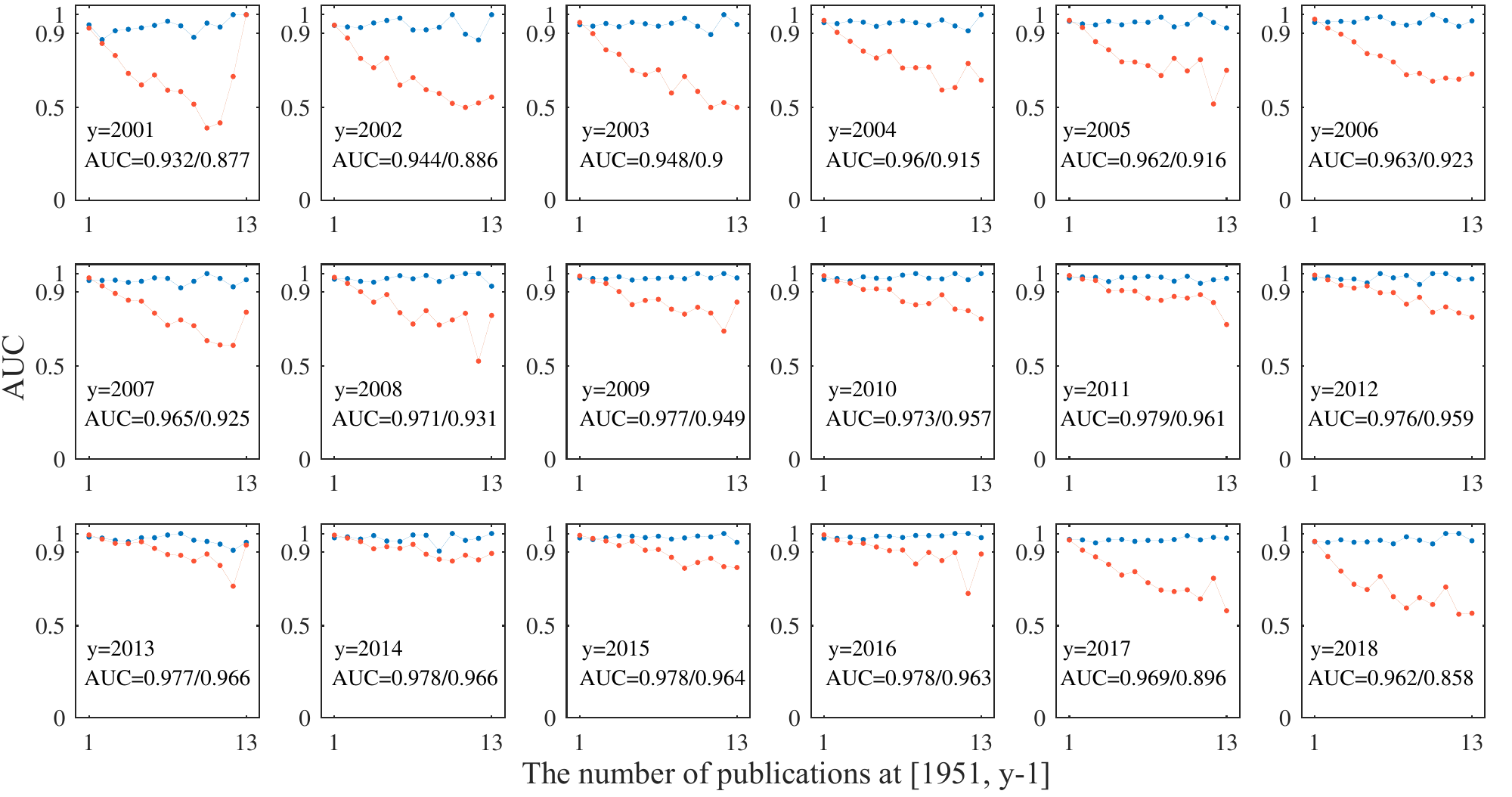}
 \caption{     {\bf  The precision of predicting the probability of publication in the following year.} The panels show the AUC of predicting the publication events in year $y$ for the tested authors who have $i$ publications within $[1951, y-1]$, where $i=1,...,13$. Red dots represent the AUC of the piecewise Poisson model, and blue dots represent that of our model. The inputs are the true numbers of publications. The AUC values are calculated based on all of the tested authors, where the value on the left of ``/" is the precision of our model and the value on the right of ``/" is that of the piecewise Poisson model. }
 \label{model_pois_annual_auc_compare}
 \end{figure*}

The distribution of the number of publications produced by authors who are partitioned into several specific subsets is a Poisson distribution, which motivated the proposal of the piecewise Poisson model\cite{ref55}.
In the piecewise Poisson model, $S_{ij}$ represents the subset of authors who produced $i$ publications within $[T_0, t_{j-1}]$, $n_{ij}$ represents the number of authors in $S_{ij}$, and $m_{ij}$ represents the sum total of the number of publications produced by those authors within $(t_{j-1}, t_j]$.
The average number of publications produced by authors in $S_{ij}$ can be obtained by calculating $m_{ij}/n_{ij}$.
The prediction formula is:
\begin{equation}\lambda_{ij}=\mathrm{e}^{\alpha_i + \beta_i (t_j - t_1)}, \label{piecewise_poisson}
\end{equation}
where $\lambda_{ij}$ is the number of publications within $(t_{j-1}, t_j]$, $i=1, ..., I$, $j=1, ..., J$. $\alpha_i$ and $\beta_i$ are calculated by using the following linear regression:
\begin{equation}
\log \frac{m_{ij}}{n_{ij}}=\alpha_i + \beta_i (t_j - t_1),
\label{piecewise_poisson_log}
\end{equation}
where $i=1, ..., I$, $j=1, ..., L$, $1<L\le{J}$.

Our model is motivated by the piecewise Poisson model. The main difference between the two models is as follows.
The expectation $x$ of Pois($x$) in Algorithm~\ref{algmodel} is a number sampled from a power-law distribution in formula~(\ref{powerlaw}).
Specifically, the expectation $x$ is dependent on the output of the LSTM with a time series as its input.
In the piecewise Poisson model, the authors are partitioned into several subsets $S_{ij}$, with each subset having equal cumulative numbers of publications of authors and different subsets having unequal numbers.
The expectation $\lambda_{ij}$ of the piecewise Poisson model is calculated by a linear regression based on the average number of publications of $S_{ij}$.
For example, authors published the same number of publications, but their time series of the number of publications can be different. Their predicted cumulative numbers of publications by the piecewise Poisson model are the same, but those by our model are different.

Here, we applied the piecewise Poisson model to the dataset.
The training dataset consists of the authors who published papers at year 2000 and their annual number of papers in years 1951--2009, involving 83,302 publications.
The test dataset consists of those the same as the training dataset and their annual number of papers in years 2010--2018, involving 137,042 publications.

The parameters are $I=40$, $J=23$, and $L=14$.
The parameter $I=40$ means that the piecewise Poisson model is not suitable for authors having published more than 40 publications in years 1951--2018.
That is, the piecewise Poisson model can be applied only to authors with no more than 13 publications in years 1951--2000.
Fig.~\ref{model_poisson_trend_compare}-\ref{model_poisson_distribution_compare} show the comparison results.
Both models perform well in predicting the publication productivity of groups over a long time interval, but poorly in predicting that of individual authors~(see indexes $s_1$ and $s_2$ in Fig.~\ref{model_poisson_trend_compare}).
The piecewise Poisson model cannot predict the publication productivity of extremely productive authors, as the fat tail does not appear in the predicted distribution~(Fig.~\ref{model_poisson_distribution_compare}).
Both of them have comparable performance in predicting the distribution of the number of publications by authors on average~(see the $p$-values shown in Fig.~\ref{model_poisson_distribution_compare}).
Our model outperforms the piecewise Poisson model in the short-term prediction of publication productivity of authors, as shown in Fig.~\ref{model_pois_annual_auc_compare}.

\subsection*{The model in reference \cite{ref56}}\label{The model in reference ref56}

 \begin{figure*}[ht]
\centering
\includegraphics[height=3.05 in,width=6 in,angle=0]{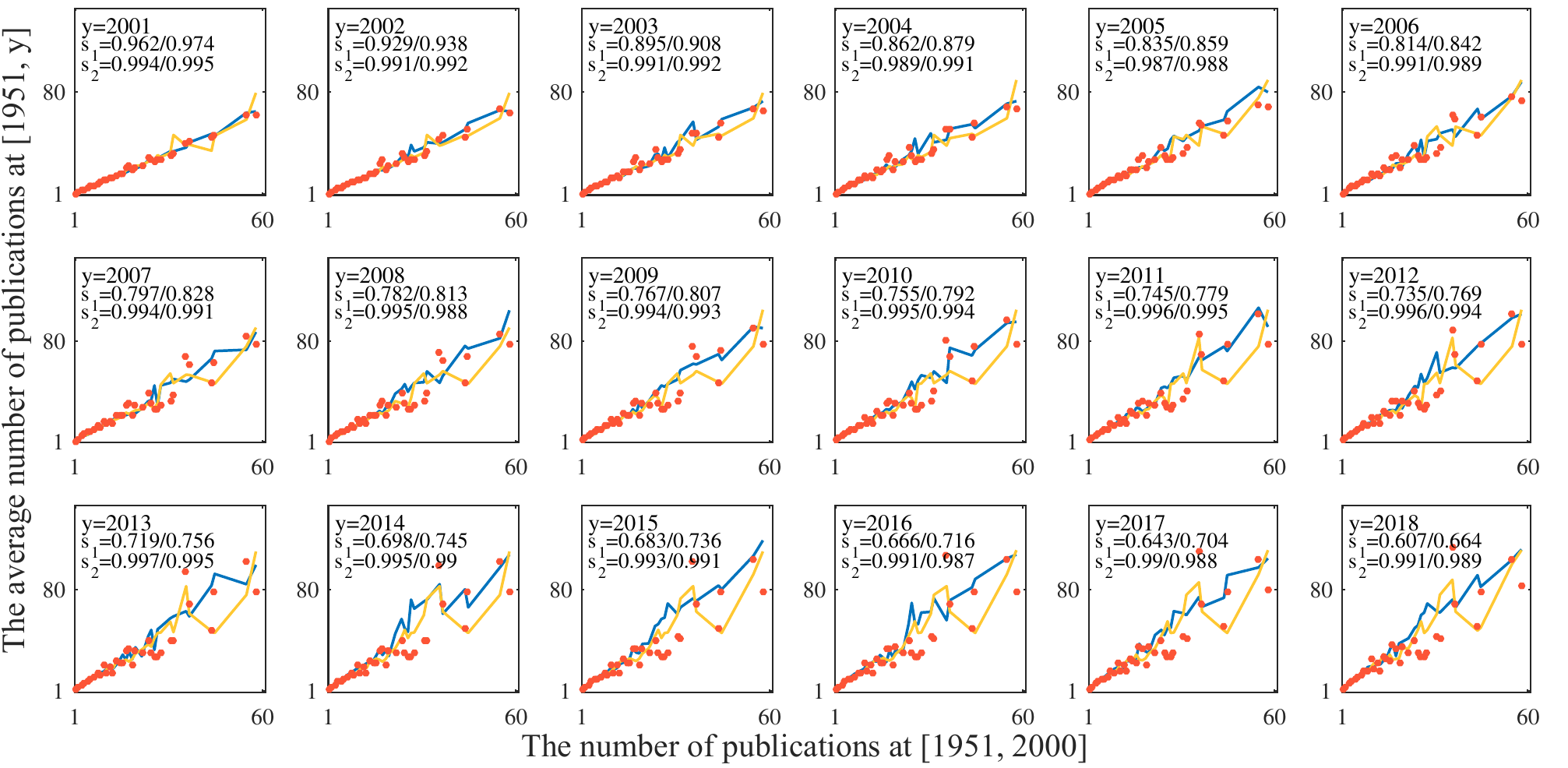}
\caption{   {\bf The trends of the number of publications.}
The tested authors are those who have $i$ publications within $[1951,2000]$, where $i=1,...,60$.
Red dots represent the average numbers of their publications within $[1951,y]$, blue lines represent the numbers predicted by the model in reference \cite{ref56}, and yellow lines represent the numbers predicted by our model.
Indexes $s_1$ and $s_2$ are the Pearson correlation coefficients defined in Fig.~\ref{model-trend-2000}, where the values on the left of ``/" are calculated by our model and the values on the right of ``/" are calculated by the model in reference \cite{ref56}.}
 \label{model_adv_trend_compare}
 \end{figure*}

  \begin{figure*}[ht]
\centering
\includegraphics[height=3.2     in,width=6.2     in,angle=0]{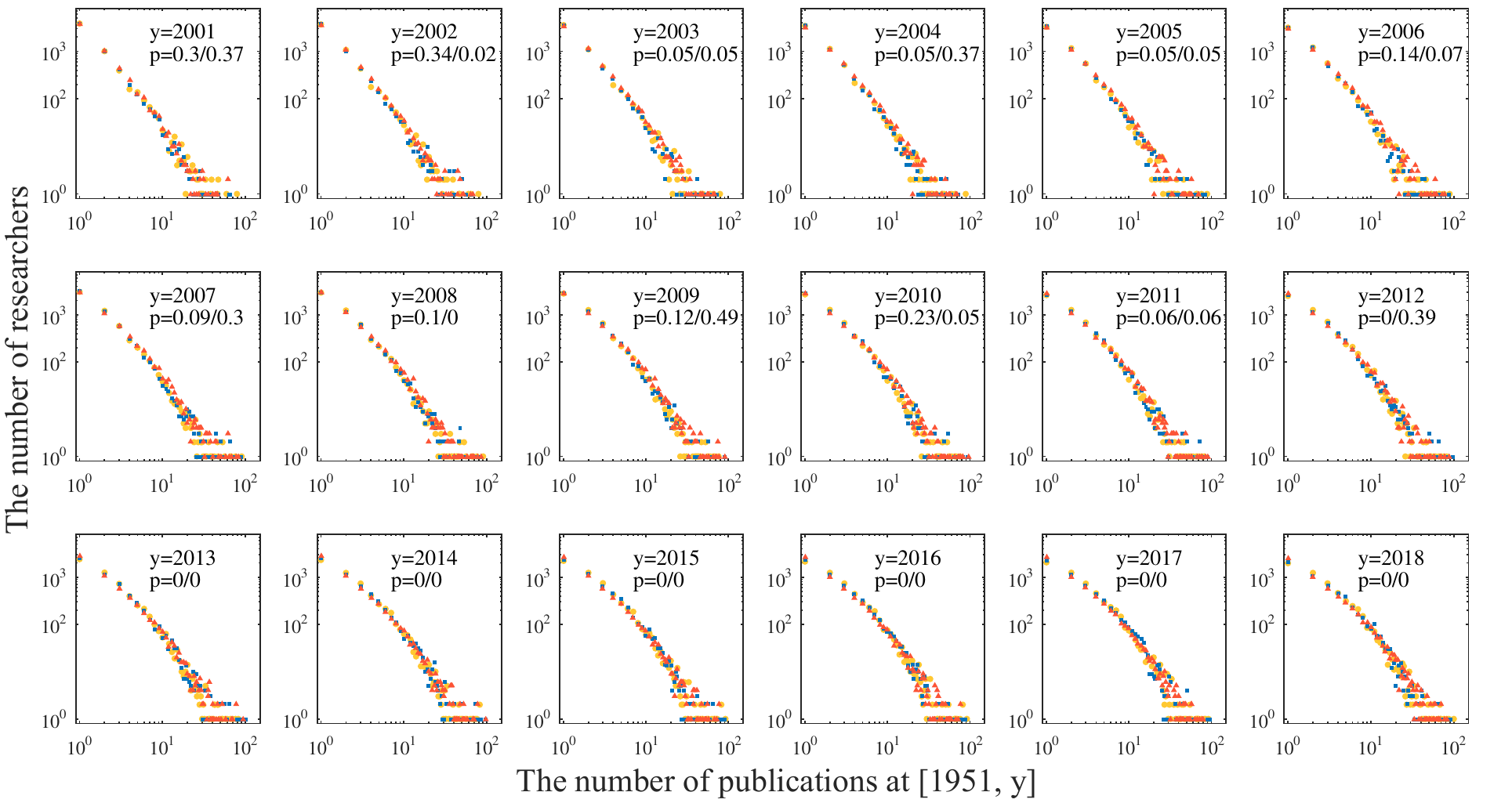}
\caption{   {\bf  The   distribution of the number of publications.} Red triangles represent the ground truth for the tested authors, blue squares represent the distributions predicted by the model in reference \cite{ref56}, and yellow circles represent the distributions predicted by our model. The $p$-value on the left of ``/" is from the KS test with the null hypothesis that the predicted distribution by our model and the true distribution are the same, and that on the right of ``/" is for the model in reference \cite{ref56}.
}
 \label{model_adv_distribution_compare}
 \end{figure*}

  \begin{figure*}[ht]
\centering
\includegraphics[height=3.35     in,width=6.2     in,angle=0]{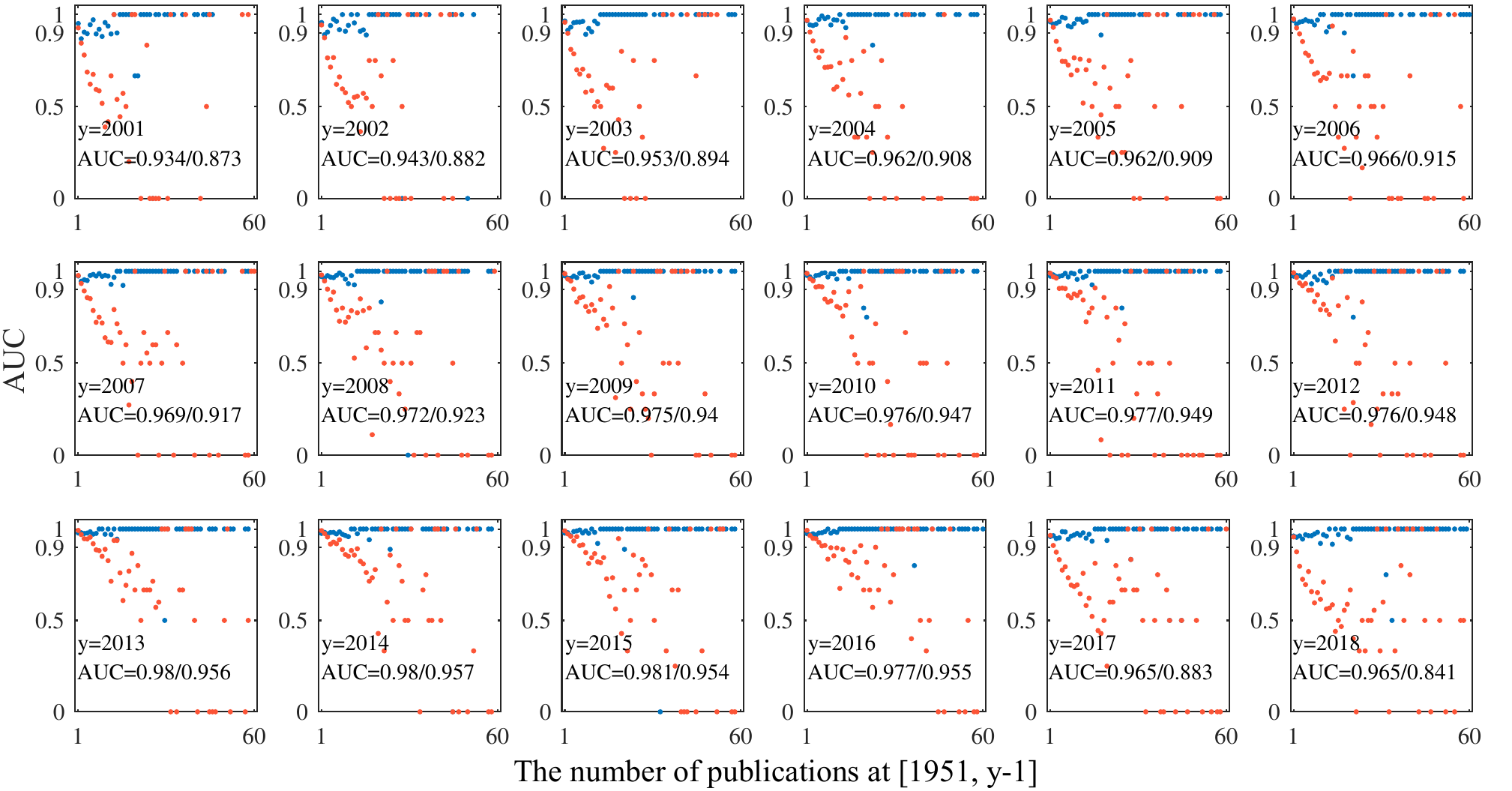}
 \caption{     {\bf  The precision of predicting the probability of publication in the following year.} The panels show the AUC of predicting the publication events in year $y$ for the tested authors who have $i$ publications within $[1951, y-1]$, where $i=1,...,60$. Red dots represent the AUC of the model in reference \cite{ref56}, and blue dots represent that of our model. The inputs are the true numbers of publications. The AUC values are calculated based on all of the tested authors, where the value on the left of ``/" is the precision of our model and the value on the right of ``/" is that of the model in reference \cite{ref56}.}
 \label{model_adv_annual_auc_compare}
 \end{figure*}

This prediction model comprises the combination of piecewise Poisson regression and log-log regression, thus it also belongs to shallow architectures.
Two prediction formulae are utilized.
One is shown as formula~(\ref{piecewise_poisson}), and the other is as follows:
\begin{equation}
\lambda_{ij}=\mathrm{e}^{\mu_j}{i^{\upsilon_j}},
\label{log_log}
\end{equation}
where $\lambda_{ij}$ is the number of publications within $(t_{j-1}, t_j]$, $m_{ij}$ and $n_{ij}$ are defined in reference \cite{ref55},  $i=1, ..., I$, $j=1, ..., J$. $\mu_j$ and $\upsilon_j$ are calculated by using the following linear regression:
\begin{equation}
\log \frac{m_{ij}}{n_{ij}}=\mu_j + \upsilon_j \log{i}.
\label{log_log_2}
\end{equation}

In this model, the training dataset is divided into 4 parts, namely, Part I, Part II, Part III, and Part IV.
Part I contains the authors in $S_{ij}$, where $i=1, ..., K$, $K<I$, $j=1, ..., L$, and $L<J$. Part II contains those in $S_{ij}$, where $i=1, ..., K$, and $j=L+1, ..., J$. Part III contains those in $S_{ij}$, where $i=K+1, ..., I$, and $j=1, ..., L$. Part IV contains those in $S_{ij}$, where $i=K+1, ..., I$, and $j=1, ..., L$.

For Parts I and II, we calculate both $\alpha_i$ and $\beta_i$ in formula~(\ref{piecewise_poisson_log}) by using linear regression, where $i=1, ..., K$, and we then obtain $\lambda_{ij}$ in formula~(\ref{piecewise_poisson}), where $j=1, ..., J$.
For Parts I and III, we calculate both $\mu_j$ and $\upsilon_j$ in formula~(\ref{log_log_2}) by using linear regression, where $i=1, ..., I$, and we then ob tain $\lambda_{ij}$ in formula~(\ref{log_log}), where $j=1, ..., L$.

For Part IV, we substitute the $\lambda_{ij}$ calculated for Part III into formula~(\ref{piecewise_poisson_log}); then, we calculate both $\alpha_i$ and $\beta_i$ by linear regression for $i=K+1, ..., I$, and finally obtain $\lambda_{ij}$ by calculating formula~(\ref{piecewise_poisson}).
$\lambda_{ij}$ in Part IV can also be obtained by formula~(\ref{log_log}), where both $\mu_j$ and $\upsilon_j$ are calculated by $\lambda_{ij}$ in Part II.
In the corresponding algorithm in reference \cite{ref56}, the average of the two values of $\lambda_{ij}$ is used as the predicted value for any author in Part IV.

This model removes a limitation of the piecewise Poisson model, which appears in the upper limit of the predicted number of publications.
We applied this model to the datasets that are the same as those applied in the piecewise Poisson model.
The parameters are $I=180$, $I_1=60$ and $K=42$.
We considered authors with no more than 60 publications in years 1951--2000.
Fig.~\ref{model_adv_trend_compare}-\ref{model_adv_annual_auc_compare} show the comparison results.
Both models have good performance in predicting the long-term publication productivity for groups but they cannot be applied to the long-term prediction for individual authors~(see indexes $s_1$ and $s_2$ in Fig.~\ref{model_adv_trend_compare}, the $p$-values in Fig.~\ref{model_adv_distribution_compare}).
All of them perform well in predicting publication events in a short time interval, and our model outperforms the model in reference \cite{ref56}~(Fig.~\ref{model_adv_annual_auc_compare}).

\subsection*{The GRU}\label{The GRU}

 \begin{figure*}[ht]
\centering
\includegraphics[height=3.2     in,width=6.2    in,angle=0]{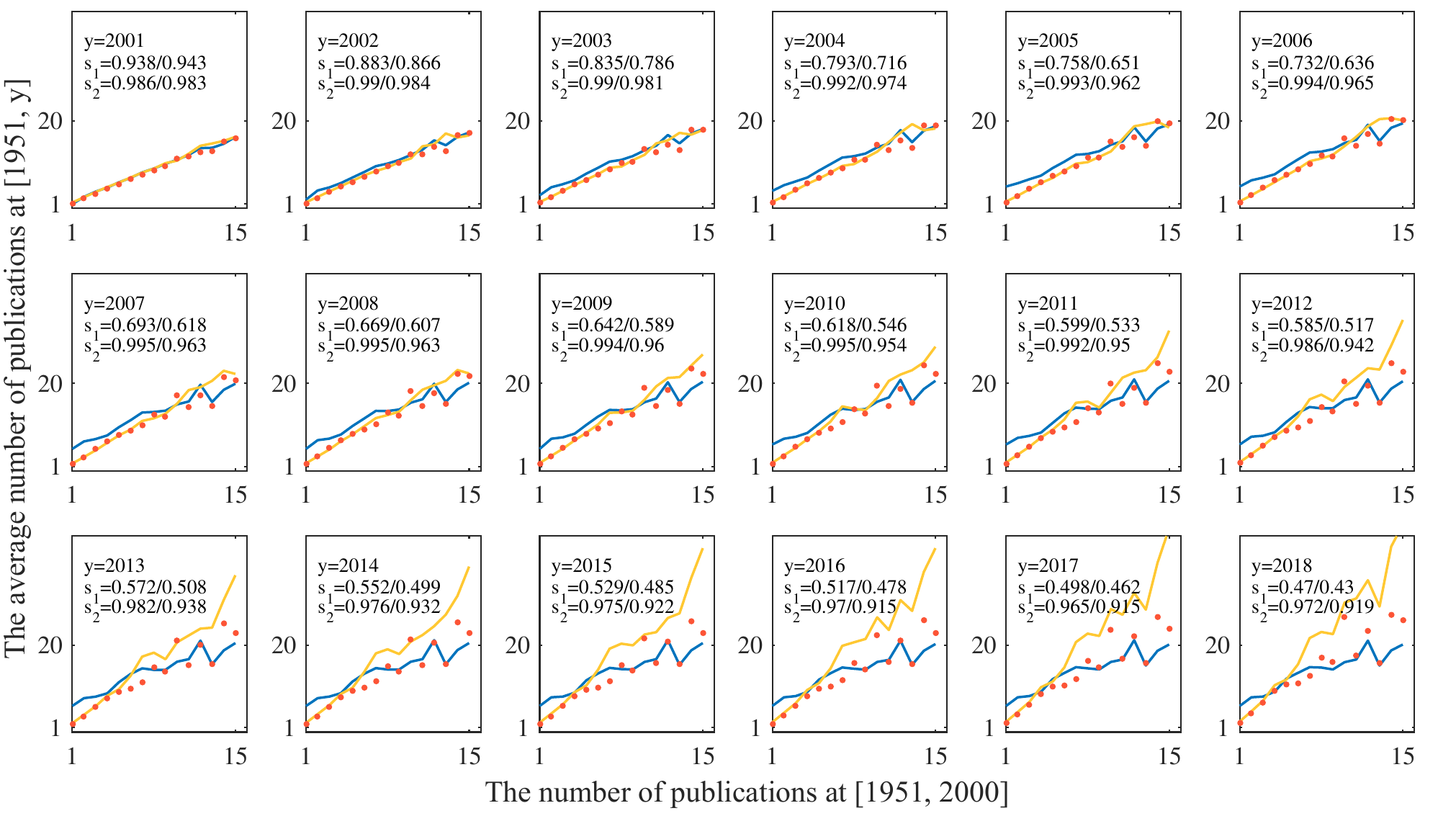}
\caption{   {\bf  The trend   of  the number of   publications.}
The tested authors are those who have $i$ publications within $[1951,2000]$, where $i=1,...,15$.
Red dots represent the average numbers of their publications within $[1951,y]$, blue lines represent the numbers predicted by the GRU, and yellow lines represent the numbers predicted by our model.
Indexes $s_1$ and $s_2$ are the Pearson correlation coefficients defined in Fig.~\ref{model-trend-2000}, where the values on the left of ``/" are calculated by our model and the values on the right of ``/" are calculated by the GRU.}
 \label{model_gru_trend_compare}
 \end{figure*}

 \begin{figure*}[ht]
\centering
\includegraphics[height=3.2    in,width=6.2     in,angle=0] {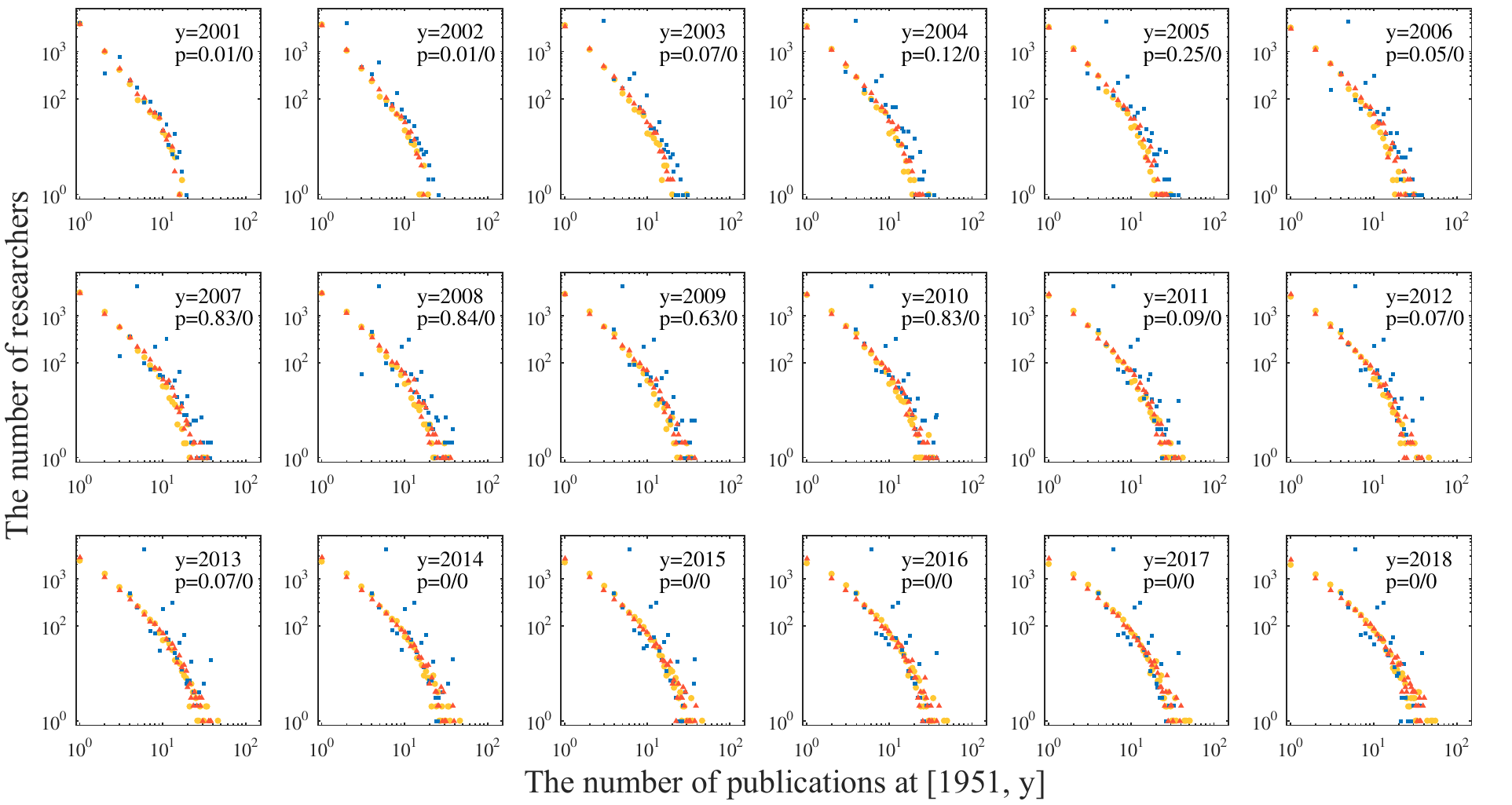}
 \caption{        {\bf  The   distribution of the number of publications.}
 Red triangles represent the ground truth for the tested authors, blue squares represent the distributions predicted by the GRU, and yellow circles represent the distributions predicted by our model. The $p$-value on the left of ``/" is from the KS test with the null hypothesis that the predicted distribution by our model and the true distribution are the same, and that on the right of ``/" is for the GRU.}
 \label{model_gru_distribution_compare}
 \end{figure*}

  \begin{figure*}[ht]
\centering
\includegraphics[height=3.2     in,width=6.2     in,angle=0]{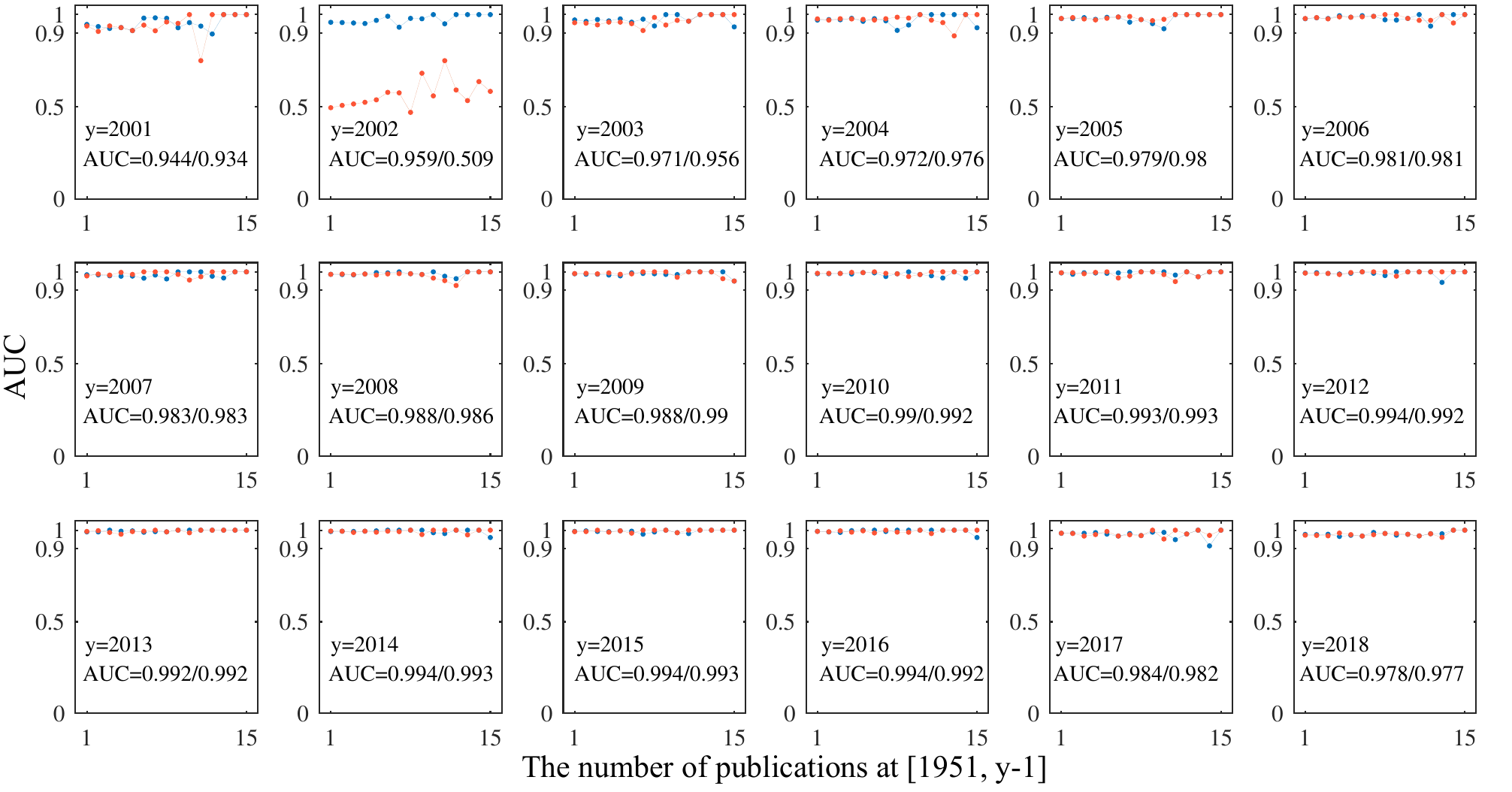}
\caption{   {\bf The precision of predicting the probability of publication in the following year.} The panels show the AUC of predicting the publication events in year $y$ for the tested authors who have $i$ publications within $[1951, y-1]$, where $i=1,...,15$. Red dots represent the AUC of the GRU, and blue dots represent that of our model. The inputs are the true numbers of publications. The AUC values are calculated based on all of the tested authors, where the value on the left of ``/" is the precision of our model and the value on the right of ``/" is that of the GRU.}
 \label{model_gru_nnual_auc_compare}
 \end{figure*}

The GRU is a gating mechanism in an RNN, which has fewer parameters than the LSTM due to the lack of an output gate.
We also used the ``Keras'' framework to design a GRU network, which possesses $32$ GRU units.
Given a time series of $12$ numerical data points, the GRU outputs one value.
A ``Dense'' layer is applied to generate one output.
ReLU is used as the activation function and MSE as the loss function.
RMSprop is chosen to optimize the network.
The training dataset used here is the same as that used in our model.
The GRU has 3,297 parameters calculated by fourfold cross validation with a time series and a target of any author in the training dataset.

The training and test datasets used here are the same as those in Section \ref{data}.
Figs.~\ref{model_gru_trend_compare}-\ref{model_gru_nnual_auc_compare} show the comparisons between the GRU and our model on the test dataset, where the input is $\{h_s(1989), h_s(1990), ..., h_s(2000)\}$ of any author $s$.
In predicting the evolutionary trend of the number of publications for groups in a long time interval, the performances of both models are comparable.
The GRU performs poorly in predicting the distribution of the number of publications, but well in predicting the publication event in a short time interval, as our model does.

\section*{Appendix C: Another example}\label{Appendix C}

\begin{figure*}[ht]
\centering
\includegraphics[height=2.3     in,width=6    in,angle=0]{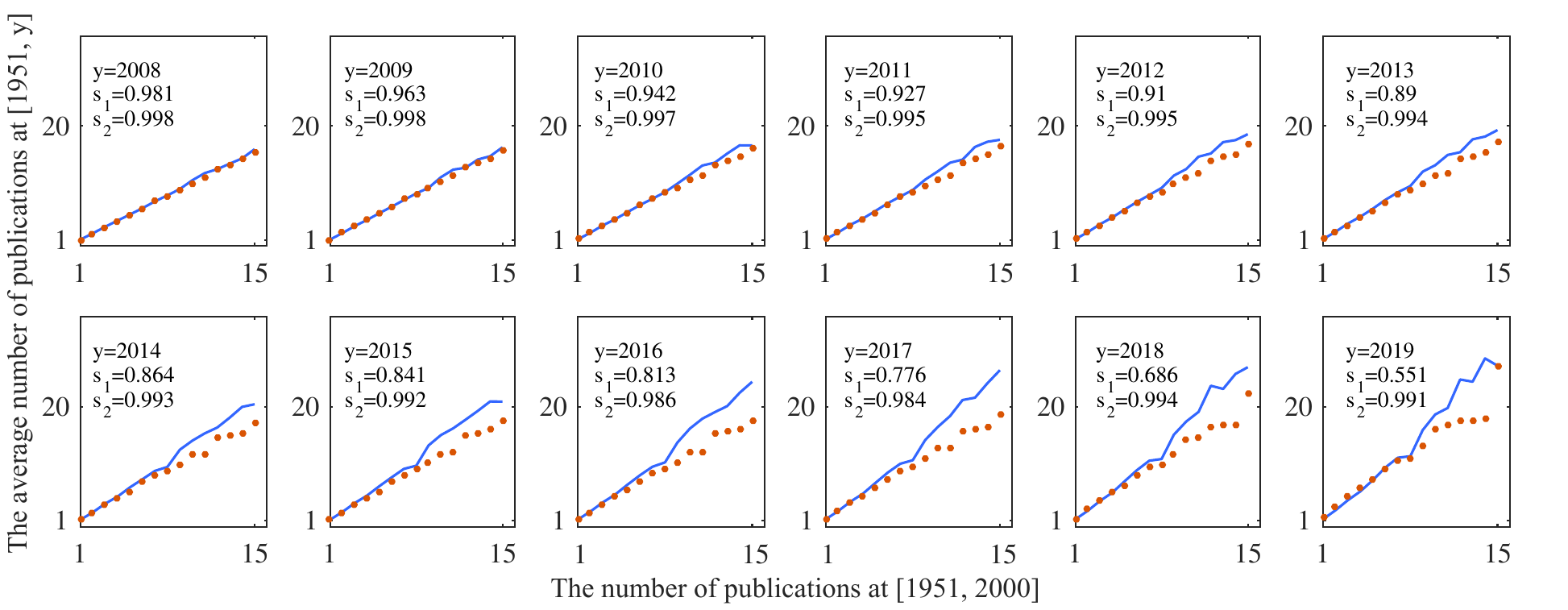}
\caption{   {\bf  The trend   of  the number of   publications predicted by the proposed model.}
The tested authors are those who have $i$ publications within $[1951,2000]$, where $i=1,...,15$.
Red dots represent the average numbers of their publications within $[1951,y]$ ($n(i,y)$), and blue lines represent the predicted numbers ($m(i,y)$).
Indexes $s_1$ and $s_2$ are the Pearson correlation coefficients defined in Fig.~\ref{model-trend-2000}.}
 \label{model-trend-another-2000}
 \end{figure*}

\begin{figure*}[ht]
\centering
\includegraphics[height=2.2     in,width=6.2     in,angle=0]{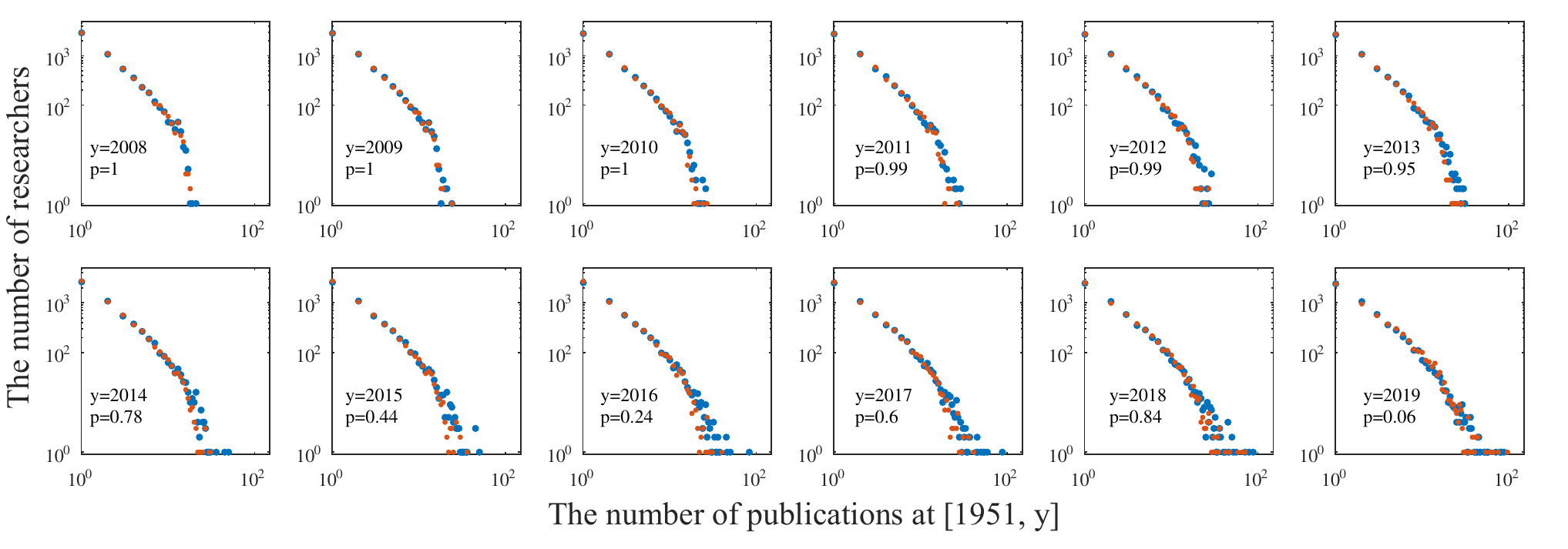}
\caption{   {\bf  The distribution of the number of publications predicted by the proposed model.} Red circles represent the ground truth for the tested authors, and blue squares represent the predicted distributions. The $p$-value is from the KS test with the null hypothesis that the compared distributions are the same.
}
 \label{model-distr-another-2000}
 \end{figure*}

\begin{figure*}[ht]
\centering
\includegraphics[height=2.2    in,width=6.2     in,angle=0]{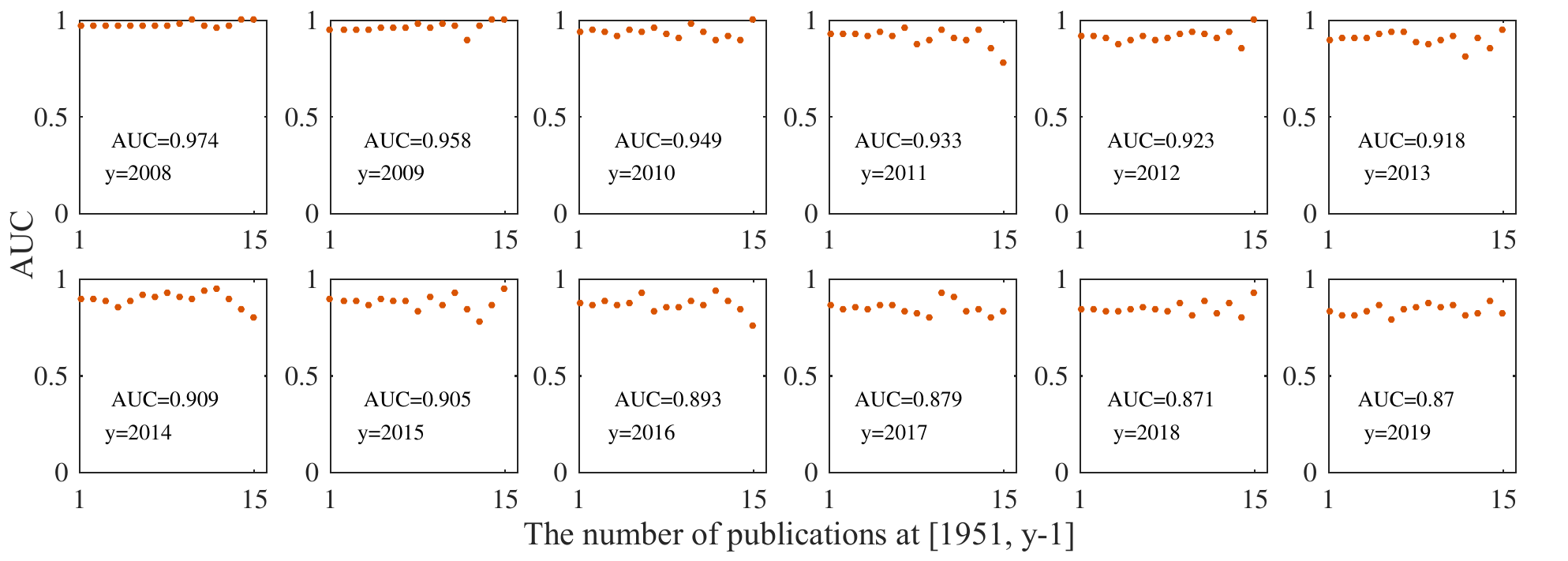}
 \caption{        {\bf  The precision of predicting publication events.} Red dots represent
the AUC of predicting the publication events in year $y$ for the tested authors who have $i$ publications within $[1951, y-1]$, where $i=1,...,15$. The inputs are the true numbers of publications.
The AUC was calculated based on all of the tested
authors.    }
 \label{model-longterm-auc-another-2000}
 \end{figure*}

The training dataset is the same as that in Section 6, but the test dataset is different. It still consists of the authors in year 2000, but their annual number of publications is from 2008 to 2019.
The parameters are $t_X=2008$ and $t_Y=2019$.
We input the time series $\{h_s(1996), h_s(1997), ..., h_s(2007)\}$ into our model to predict the number of publications in the following years.
Figs.~\ref{model-trend-another-2000}-\ref{model-longterm-auc-another-2000} show the results of our model.

\begin{thebibliography}{1}

\bibitem{Garca-Otero2019}
Garc\'ia-Suaza A, Otero  J,   Winkelmann  R  (2019)  Predicting early career productivity of PhD economists: Does advisor-match matter?  Scientometrics, 122, 429-449.

\bibitem{Clauset-Larremore2017}
Clauset  A, Larremore  DB,   Sinatra  R  (2017)  Data-driven predictions in the science of science. Science, 355(6324), 477-480.

\bibitem{Sinatra2016}
 Sinatra  R, Wang  D, Deville  P, Song  C,   Barab\'asi  AL  (2016) Quantifying the evolution of individual scientific impact. Science 354, aaf5239.

  \bibitem{Lindahl-Colliander2019}
  Lindahl  J, Colliander  C,   Danell  R  (2020)  Early career performance and its correlation with gender and publication output during doctoral education. Scientometrics,  122(1), 309-330.

    \bibitem{Ejermo-Fassio2019}
Ejermo O, Fassio  C,   K\"allstr\"om  J  (2019)  Does mobility across universities raise scientific productivity?
 Oxford B Econ Stat,  82(3), 603-624.

 \bibitem   {XieLL2018}
Xie  Z, Li  M, Li  JP, Duan  XJ,   Ouyang  ZZ  (2018)  Feature analysis of multidisciplinary scientific collaboration patterns based on PNAS. EPJ Data Science,  7, 1-17.

\bibitem
{Schmidhuber2015}
Schmidhuber J (2015) Deep learning in neural networks: An overview. Neural Networks, 61, 85-117.


\bibitem
{lstm1997}
Hochreiter S, Schmidhuber J  (1997) Long short-term memory. Neural Comput, 9(8), 1735-1780.

\bibitem{Nelder-Wedderburn1972}
Nelder  JA,   Wedderburn  RW  (1972) Generalized linear models. J R Stat Soc Ser A-G, 135(3), 370-384. 

\bibitem
{ref55}
Xie Z (2020) Predicting publication productivity for researchers: A piecewise poisson model. J  Informetr, 14(3), 101065.

\bibitem
{ref56}
Xie Z (2020) A prediction method of publication productivity for researchers. IEEE Trans Comput Soc Syst. (to be accepted)


\bibitem
{Garfield1955}
Garfield E (1955) Citation indexes for science: A new dimension in documentation through association of ideas. Science, 122 (3159), 108-111.

\bibitem
{Harnad2009}
Harnad S (2009) Open access scientometrics and the UK research assessment exercise. Scientometrics, 79(1), 147-156.

\bibitem{Hirsch2005}
Hirsch  JE  (2005)  An index to quantify an individual's scientific research output. Proc  Natl  Acad  Sci  USA,  102, 16569-16572.

    \bibitem{Way-Morgan2019}
Way  SF, Morgan  AC, Larremore  DB,   Clauset  A  (2019)  Productivity, prominence, and the effects of academic environment. Proc Natl Acad Sci USA, 116(22), 10729-10733.

\bibitem{Mazloumian2012}
 Mazloumian  A (2012)  Predicting researchers' scientific impact. Plos One, 7(11), 1-5.

\bibitem{Wang2013}
 Wang  D, Song C,   Barab\'asi AL  (2013)  Quantifying long-term scientific impact. Science, 342(6154), 127-132.

\bibitem{Cao2016}
 Cao  X, Chen  Y,   Liu  KR  (2016)  A data analytic approach to quantifying scientific impact. J  Informetr, 10(2), 471-484.

 \bibitem{Newman2014}
Newman  MEJ  (2014)  Prediction of highly cited papers.  Europhys Lett, 105(2), 28002. 

\bibitem{Pobiedina2016}
Pobiedina  N,   Ichise  R  (2016)  Citation count prediction as a link prediction problem. Appl Intell, 44(2), 252-268.

\bibitem{Klimek2016}
Klimek  P, Jovanovic  AS, Egloff  R,   Schneider  R  (2016)  Successful fish go with the flow: Citation impact prediction based on centrality measures for term-document networks. Scientometrics, 107(3), 1265-1282.


\bibitem{Stern2014}
Stern  DI (2014)  High-ranked social science journal articles can be identified from early citation information. Plos One, 9(11), e112520.

\bibitem{Abramo2019}
Abramo  G, D'Angelo  CA,   Felici  G  (2019)  Predicting publication long-term impact through a combination of early citations and journal impact factor. J  Informetr, 13(1), 32-49.


\bibitem{Kosteas2018}
 Kosteas  VD (2018)  Predicting long-run citation counts for articles in top economics journals. Scientometrics, 115(3), 1395-1412.


\bibitem{Bornmann2014}
Bornmann  L, Leydesdorff  L,   Wang  J  (2014)  How to improve the prediction based on citation impact percentiles for years shortly after the publication date?  J  Informetr, 8(1), 175-180.

\bibitem
{HuTai2020}
Hu YH, Tai CT, Liu KE, Cai CF (2020) Identification of highly-cited papers using topic-model-based and bibliometric features: The consideration of keyword popularity. J Informetr, 14, 101004.

\bibitem{Bai2019}
  Bai XM,  Zhang LI,  Lee I (2019)  Predicting the citations of scholarly paper, J  Informetr, 13,  407-418.

\bibitem{YuYu2014}
Yu  T, Yu  G, Li  PY,   Wang  L  (2014)  Citation impact prediction for scientific papers using stepwise regression analysis. Scientometrics, 101(2),
1233-1252.

\bibitem{Abrishami2019}
Abrishami  A,   Aliakbary  S  (2019)  Predicting citation counts based on deep neural network learning techniques. J Informetr, 13(2), 485-499.


\bibitem
{WangFan2019}
Wang FH, Fan Y, Zeng A, Di ZR (2019) Can we predict ESI highly cited publications? Scientometrics, 118, 109-125.


\bibitem
{RuanZhu2020}
Ruan XM, Zhu YY, Li J, Cheng Y (2020) Predicting the citation counts of individual papers via a BPneural network. J Informetr, 14, 101039.

\bibitem
{XuLi2019}
Xu JG, Li MJ, Jiang J, Ge BF, Cai MS (2019) Early prediction of scientific impact based on multi-bibliographic features and convolutional neural network. IEEE Access, 7,  92248-92258.

\bibitem
{Mistele2019}
Mistele T, Price T, Hossenfelder S (2019) Predicting authors' citation counts and $h$-indices with a neural network. Scientometrics, 120, 87-104.

\bibitem
{Ye2008}
Ye Fred Y, Rousseauc R (2008) The power law model and total career $h$-index sequences. J  Informetr, 2(4), 288-297.

\bibitem
{Egghe2006}
Egghe L, Rousseau R  (2006) An informetric model for the hirsch-index. Scientometrics, 69(1), 121-129.


 \bibitem{Acuna2012}
Acuna  DE, Allesina  S,   Kording  KP  (2012)  Future impact: Predicting scientific success. Nature, 489(7415), 201.


 \bibitem{Dong2016}
Dong  Y, Johnson  RA,  Chawla  NV  (2016)  Can scientific impact be predicted?  IEEE Transactions on Big Data, 2(1), 18-30.


 \bibitem{Laurance2013}
Laurance  WF, Useche DC, Laurance  SG,   Bradshaw  CJ (2013) Predicting publication success for biologists. BioScience, 63(10), 817-823.


 \bibitem{Lehman2017}
Lehman  HC  (2017)  Age and achievement (Vol. 4970). Princeton University Press.


 \bibitem{Simonton1984}
Simonton  DK  (1984) Creative productivity and age: A mathematical model based on a two-step cognitive process. Dev  Rev, 4(1), 77-111.


 \bibitem{Way-Morgan2017}
Way  SF, Morgan  AC, Clauset  A,   Larremore  DB  (2017)  The misleading narrative of the canonical faculty productivity trajectory. Proc Natl Acad
Sci USA, 114(44), 9216-9223.


\bibitem{Price1965}
Price DJS (1965) Networks of scientific papers. Science, 149(3683), 510-515.


\bibitem
{Consul1973}
Consul PC, Jain GC (1973) A generalization of the Poisson distribution.
Technometrics, 15(4), 791-799.


\bibitem
{XieOu2016}
Xie Z, Ouyang ZZ, Li JP (2016) A geometric graph model for coauthorship networks. J Informetr, 10, 299-311.


\bibitem
{XieOu2018}
Xie Z, Ouyang ZZ, Li JP, Dong EM, Yi DY (2018) Modelling transition phenomena of scientific coauthorship networks. J Assoc Inf Sci Tech, 69(2), 305-317.


\bibitem
{Xie2019a}
Xie Z (2019) A cooperative game model for the multimodality of coauthorship networks. Scientometrics, 121(1), 503-519.

\bibitem
{ReLU2010}
Nair V, Hinton GE (2010) Rectified Linear Units Improve Restricted Boltzmann Machines Vinod Nair. In ICML, 807-814.


\bibitem{Hollander}
Hollander  M, Wolfe  DA  (1973) Nonparametric statistical methods. Wiley.


\bibitem{Xie2020coauthors}
Xie Z  (2020) Predicting the number of coauthors for researchers: A learning model. J  Informetr, 14(2), 101036.


\end{thebibliography}
\end{document}